\documentstyle[prl,aps,preprint,tighten,floats,aps,epsf,psfig]{revtex}

\def\be{\begin{equation}}
\def\ee{\end{equation}}
\def\beq{\begin{equation} }
\def\eeq{\end{equation} }
\def\bear{\begin{eqnarray}}
\def\eear{\end{eqnarray}}
\def\beqn{\begin{eqnarray}}
\def\eeqn{\end{eqnarray}}
\def\MK{M}
\def\MP{M_{Pl}}
\def\MS{M_{string}}
\def\IS{I^{singlets}}
\def\AS{A^{singlets}}
\begin{document}
\draft
\preprint{UPR-0750-T, IEM-FT-155/97, hep-ph/9705391}
\date{\today}
\title{Intermediate Scales, $\mu$ Parameter, and Fermion Masses from String
Models} 
\author{Gerald Cleaver, Mirjam Cveti\v c, Jose R. Espinosa, Lisa Everett and
Paul Langacker
\
}
\address{Department of Physics and Astronomy \\ 
          University of Pennsylvania, Philadelphia PA 19104-6396\\ }
\maketitle
\begin{abstract}
We address intermediate scales within a class of string models. 
The intermediate scales occur due to the  SM singlets
$S_i$ acquiring non-zero VEVs due to radiative
breaking; the mass-square  $m_i^2$ of $S_i$ is driven negative  at
$\mu_{ RAD}$ due to ${\cal O}(1)$ Yukawa couplings of $S_i$ to exotic 
particles (calculable  in a class of string models). The actual  VEV of $S_i$ 
depends on the relative magnitude of the non-renormalizable terms of the type 
$\hat{S}_i^{K+3}/M^K$ in the superpotential. We mainly consider the 
case in which the $S_i$ are charged under an additional non-anomalous $U(1)'$ 
gauge symmetry and
the VEVs occur along $F$- and $D$-flat directions. We explore various  
scenarios in detail, depending on the type of Yukawa couplings to the 
exotic particles and
on the initial boundary values of the soft SUSY breaking parameters. We then
address the implications of these scenarios for the $\mu$ parameter and the
fermionic masses of the standard model.
\end{abstract}
\pacs{}

\section{Introduction}
One  prediction of 
the  weakly coupled heterotic  string  is the tree level 
 gauge coupling unification at 
$\MS \sim g_U \times 5 \times 10^{17} $ GeV \cite{K},
 where $g_U$ is the gauge coupling at the string scale. $\MS$ is the
only mass scale  that  appears in the effective Lagrangian of such string
vacua, and thus is one mass scale naturally provided by string theory.

However, one of the major obstacles to connecting string theory to the 
low energy   world is the absence of a fully satisfactory
 scenario for supersymmetry (SUSY)  breaking, either at the level
 of world-sheet dynamics or at the level of the effective theory. 
The SUSY breaking induces soft mass parameters which
provide another scale in the theory that can hopefully 
provide a link between  $\MS$ and $M_Z$, the scale of
electroweak symmetry breaking. For example, in models with radiative breaking one
of the Higgs mass-squares runs from an initial positive value $m_0^2$ at
$\MS$ to a negative value, of ${\cal O}(-m_0^2)$, at low energies, so that
the electroweak scale is set by the soft supersymmetry breaking scale
$m_0$ (and
not by the intermediate scale at which the mass-square goes through zero).

In spite of this difficulty,
string theory does provide certain generic and, 
 for a  certain class of string vacua, definite predictions. 
With the assumption of 
soft supersymmetry breaking masses  as free parameters, the 
features of the string models, such as the explicitly calculable structure
of
the superpotential, provide  specific
predictions for the  low energy physics. 

For example, one can restrict the analysis to a set of string vacua
which have $N=1$ supersymmetry,
  the standard  model (SM) gauge group as a part of the
 gauge structure, and a particle content that includes
 three SM families  and  at least two SM
Higgs doublets, {\it i.e.},  the string vacua which
have at least the ingredients of the MSSM and thus the potential to be
realistic\footnote{
A number of such models
(not necessarily consistent with gauge unification)
were constructed
as orbifold models\cite{DHVW,FIQS} with  Wilson
 lines, as well as models based on the free (world-sheet)
 fermionic constructions\cite{ABK,NAHE,faraggi90a,CHL}. 
For review and references see\cite{dienes}.}.  Such
vacua often
predict an additional nonanomalous $U(1)'$ gauge symmetry in the observable
sector.
It has been argued \cite{CL} that for this class of string
vacua with an additional $U(1)'$ broken by a single standard model singlet $S$,
the mass scale of the $U(1)'$ breaking should be in the electroweak range (and not
larger than a TeV).  
That is,
if the $U(1)'$ is not broken at a large scale through string dynamics,
the $U(1)'$ breaking may be radiative if there are Yukawa couplings of
${\cal O}(1)$ of $S$ to exotic particles.  The scale
of the symmetry breaking is then set by the soft supersymmetry breaking
scale $m_0$, in analogy to the radiative breaking of the electroweak
symmetry described above.

Recently, a model was considered \cite{CDEEL} in which the two SM Higgs
doublets couple to the SM singlet, and the gauge symmetry breaking
scenarios and mass spectrum were analyzed in detail. A major conclusion of
this analysis was that a large class of string models not only predicts the
existence of additional gauge bosons and exotic matter particles, but can
often ensure that their masses are in the electroweak range.
Depending on the values of the assumed soft supersymmetry breaking
mass parameters at $\MS$, each specific model leads to calculable
predictions, which can satisfy the phenomenological bounds.  In addition,
the model considered in \cite{CDEEL,sykm} forbids an elementary $\mu$ term for
appropriate $U(1)'$ charges, but an effective $\mu$ is generated by the
electroweak scale VEV of the singlet, thus providing a natural solution to
the $\mu$ problem.

However, the qualitative picture changes if there are couplings in the
renormalizable superpotential of exotic particles to two or more
mirrorlike singlets $S_i$ charged under the $U(1)'$.  In this case, the
potential may have $D$ and $F$-flat directions, along which it consists only of
the quadratic mass terms due to the soft supersymmetry breaking mass squared
parameters $m_i^2$. If there is a mechanism to drive the linear combination $m^2$ 
that is relevant along the flat directions negative at
$\mu_{ RAD}\gg M_{Z}$, the $U(1)'$ breaking is at an intermediate scale. On the
other hand, if some individual $m_i^2$ are negative but $m^2$ remains positive,
then the $D$-flat direction is not relevant and the breaking occurs near the
electroweak scale, similar to the case of only one singlet. 

A large number of string models have the ingredients that can lead to such
scenarios:
\begin{itemize}
\item
SM singlets $S_i$ which do not have renormalizable self-interactions of
the superpotential ($F$-flatness).
\item
If such  singlets $S_i$ are charged under additional nonanomalous
$U(1)'$ factors, more than one $S_i$ with opposite relative signs for
the additional $U(1)'$ charges may ensure $D$-flat directions. This is the case
that we focus on in this paper. However, similar
considerations hold for a single scalar $S$ which carries no gauge quantum numbers
and therefore has no $D$-terms.
\item
Most importantly, in a large class of models such $S_i$ can couple to
additional
exotic particles via Yukawa couplings  of ${\cal O}(1)$.
Such Yukawa couplings can then lead to radiative breaking, by driving
some or all of the soft $m_i^2$ parameters negative at $\mu_{ RAD}\gg M_Z$.
\end{itemize}

In the case of pure radiative breaking, the minimum of the potential
occurs near the scale $\mu_{ RAD}$, and so the nonzero VEV of $S_i$'s is at an
intermediate scale.   In principle, non-renormalizable terms in the
superpotential compete with the radiative breaking.  These terms are
generically present in most string models.
If such
terms dominate at scales below $\mu_{ RAD}$, they will determine the VEV of
$S_i$.  In this case, the order of magnitude of the VEV depends on the
order of the
non-renormalizable terms, but is also at an intermediate
scale.

The purpose of this paper is to investigate the nature of intermediate scales in a
class of string models.  Intermediate scales are of importance, as they are often
utilized in phenomenological models (e.g., for neutrino masses), and may also have
important cosmological implications
(e.g., in the inflationary scenarios
\cite{infl}). In this paper, we also investigate the implications of intermediate
scales for the standard model sector of the theory, specifically for the $\mu$
parameter, ordinary fermion masses, and Majorana and Dirac neutrino masses. 
 
In Section II, we give a general discussion of radiative breaking along a flat
direction and study two different mechanisms (radiative corrections and
non-renormalizable terms) that stabilize the potential and fix an intermediate
scale VEV. We also examine the implications for the low energy particle spectrum
of such type of scenarios.

In Section III, we explore the range of $\mu_{RAD}$ that can arise 
assuming that the flat direction has large Yukawa couplings to exotic fields (as
is typically expected in string models) \cite{BL}. We consider three different models,
with varied quantum numbers for the exotic fields, and in each case we examine
the effect on $\mu_{ RAD}$ of different choices of boundary conditions for the
soft masses. The relevant renormalization group equations, with exact analytic
solutions and useful simplified approximations are given in Appendix~A.

In Section IV, we discuss the size and structure of non-renormalizable
contributions \cite{binetruy96}
to the superpotential expected in string models
\cite{nan90,F,faraggi92}. These terms
are relevant to fix the intermediate scale and can also play an important
role in connection with the physics of the effective low-energy 
theory. 
In particular, in Section~V we study how these contributions may offer a 
natural
solution to the $\mu$ problem and generate 
a hierarchy of standard model ordinary fermion masses in rough agreement
with observation. 
We also indicate that interesting neutrino masses
can arise from such terms. Both the ordinary seesaw mechanism for Majorana masses,
or naturally small (non-seesaw) Dirac or Majorana masses can be generated. 

Finally, in Section VI we draw some conclusions.

\section{Intermediate Scale VEV}

A well known mechanism to generate intermediate scale VEVs in supersymmetric
theories utilizes the flat directions generically present in these models 
\cite{fabio}. The discussion in this section applies to a general class of
supersymmetric models with flat directions; string models \cite{flatorb}
discussed subsequently in general possess these features.

For example, consider a model with two chiral multiplets $\hat{S}_1$ and
$\hat{S}_2$ that are singlets under the standard model gauge group, but carry
charges $Q_1$ and $Q_2$ under an extra 
$U(1)'$\footnote{We assume that the supersymmetry breaking is due to hidden 
sector fields 
that are not charged under the additional $U(1)'$, i.e., the $U(1)'$ 
 belongs to the observable sector. Thus,  the  mixing of the $U(1)_Y$ and 
$U(1)'$  gauge kinetic 
energy terms, which can arise due to the 
one-loop   
(field theoretical) corrections or genus-one corrections in string 
theory \cite{mixdan}, can be neglected in the analysis of the soft
supersymmetry breaking 
mass parameters.}.
If these charges have opposite signs ($Q_1Q_2<0$), the scalar field
direction $S$ with
\begin{equation}
\langle S_1\rangle=\cos\alpha_Q\langle S\rangle,\;\;\;\;
\langle S_2\rangle=\sin\alpha_Q\langle S\rangle,
\label{flatdir}
\end{equation}
with
\begin{equation}
\tan^2\alpha_Q\equiv
\frac{|Q_1|}{|Q_2|},
\end{equation}
is $D$-flat. If $\hat{S}_1$ and $\hat{S}_2$ do not couple among themselves
in the renormalizable superpotential the direction
(\ref{flatdir}) is also $F$-flat and the only contribution to the scalar potential
along $S$ is given by the soft mass terms $m_1^2|S_1|^2+m_2^2|S_2|^2$. If we
concentrate on the (real) component $s=\sqrt{2} {\mathrm Re} S$ along the flat
direction:
\begin{equation}
\label{flatreal}
s=s_1 \cos\alpha_Q  + s_2 \sin\alpha_Q ,
\end{equation}
the potential is simply
\begin{equation}
V(s)=\frac{1}{2}m^2s^2,
\end{equation}
where
\begin{equation}
\label{msum}
m^2=m_1^2\cos^2\alpha_Q+m_2^2\sin^2\alpha_Q=\left(
\frac{m_1^2}{|Q_1|}+\frac{m_2^2}{|Q_2|}\right)
\frac{|Q_1Q_2|}{|Q_1|+|Q_2|},
\end{equation}
which is evaluated at the scale $\mu=s$. We assume that
$m^2$ is positive at the 
string scale\footnote{In some string models it is in principle possible to  
obtain $m^2<0$ for some scalar field, depending on its modular weight
\cite{madrid}.} 
($m^2=m_0^2$ if we assume universality.) However, $m^2$ can
be driven to negative values at the electroweak scale if $\hat{S}_1$ and/or
$\hat{S}_2$ have a large
Yukawa coupling  to other fields in the superpotential\footnote{Another case which
often
occurs is that in which, e.g., $m_1^2$ goes negative but $m^2$ remains positive. 
In that case the $D$-flatness is not important: $S_1$ acquires an electroweak
scale 
VEV while $\langle S_2\rangle=0$, so that the $U(1)'$ is broken at or near the
electroweak scale, 
similar to the case discussed in \cite{CL,CDEEL}.}. In this case, the
potential develops
a minimum along the flat direction and $S$ acquires a VEV. From the
minimization condition
\begin{equation}
\frac{dV}{ds}=\left.\left(m^2+\frac{1}{2}\beta_{m^2}\right)\right|_{\mu=s}s=0,
\end{equation}
(where $\beta_{m^2}=\mu\frac{dm^2}{d\mu}$) one sees that the VEV $\langle
s\rangle$ is determined by \begin{equation}
m^2(\mu=\langle s\rangle)=-\frac{1}{2}\beta_{m^2},
\end{equation}
which is satisfied very close to the scale $\mu_{ RAD}$ at which $m^2$ crosses
zero. This scale is fixed by the renormalization group evolution of
parameters from $\MS$ down to the electroweak scale and will lie
at some intermediate scale. The precise value depends on the
couplings of $\hat{S}_{1,2}$ and the particle content of the model, as we discuss
in the next section.
 
The stabilization of the minimum along the flat direction can also be
due to non-renormalizable terms in the superpotential, which lift the
flat direction for sufficiently large values of $s$. If these terms are 
important
below the scale $\mu_{ RAD}$, they will determine $\langle s\rangle$.
The relevant non-renormalizable terms\footnote{The notation for superfields 
and their bosonic and fermionic components follows that of \cite{CDEEL}.}
are of the form\footnote{One can also 
have terms of the form
$\alpha_{K}^{K}\hat{S}^{2+K}\hat{\Phi}/{\MP^K}$,
where $\Phi$ is a standard model singlet that does not acquire a VEV.
These have similar implications as the terms in (\ref{nrsup}).}
\begin{equation}
\label{nrsup}
W_{\rm NR}=\left(\frac{\alpha_{K}}{\MP}\right)^{K}\hat{S}^{3+K},
\end{equation}
where $K=1,2...$
and $\MP$ is the Planck scale. The coefficients $\alpha_{K}$
will be discussed in Section IV. 
Depending on the $U(1)'$ charges, not all values of
$K$ are allowed. For example, if $Q_1=-Q_2$, $U(1)'$ invariance dictates
$W_{\rm NR}\sim (\hat{S}_1\hat{S}_2)^n\sim \hat{S}^{2n}$ and only odd values of $K$
should be considered. If $Q_1=\frac{4}{5}$, $Q_2=\frac{-1}{5}$, $W_{\rm NR}\sim
(\hat{S}_1\hat{S}_2^4)^n\sim \hat{S}^{5n}$, and so on. 

Including the $F$-term from (\ref{nrsup}), the potential along $s$ is
\begin{equation}
\label{potflat}
V(s)= \frac{1}{2}m^2s^2+\frac{1}{2(K+2)}
\left(\frac{s^{2+K}}{{\cal M}^K}\right)^2,
\end{equation}
where ${\cal M} = {\cal C}_K \MP/ \alpha_K$, and the coefficient
${\cal C}_K = [ 2^{K+1}/((K+2)(K+3)^2)]^{1/(2K)}$
takes the values $(0.29,\, 0.53,\, 0.67,\, 0.76,\, 0.82)$ 
for $K= (1,2,3,4,5)$.
The VEV of $s$ is then\footnote{For simplicity, we do not 
include
in (\ref{potflat}) soft-terms of the type $(AW_{\rm NR}+{\mathrm H.c.})$ 
with $A\sim
m_{soft}$. Such terms do not affect the order of
magnitude estimates that follow.} 
\begin{equation}
\label{vev}
\langle s\rangle=
\left[\sqrt{(-m^2)}{\cal M}^K \right]^\frac{1}{K+1}
= \mu_K\sim (m_{soft}{\cal M}^K)^\frac{1}{K+1},
\end{equation}
where $m_{soft}={\cal O} (|m|)={\cal O} (M_Z)$ is a typical soft supersymmetry
breaking scale.
In this equation, $-m^2$ is evaluated at the scale $\mu_K=\langle s\rangle$
and has to satisfy the necessary condition $m^2(\mu_K)<0$. 
If non-renormalizable terms are negligible below $\mu_{ RAD}$, no solution
to (\ref{vev}) exists and $\langle s\rangle$ is fixed solely by the running $m^2$.

The mass $M_S$ of the physical field $s$ in the vacuum $\langle s\rangle$
can be obtained easily in both types of breaking scenarios. In both cases, $M_S$
is of the soft breaking scale or smaller and not of the intermediate scale
$\langle s\rangle$. For pure radiative breaking,
\begin{equation}
M_S^2\equiv\left.\frac{d^2V}{ds^2}\right|_{s=\langle s \rangle}=
\left.\left(\beta_{m^2}+\frac{1}{2}\mu\frac{d}{d\mu}\beta_{m^2}\right)
\right|_{\mu=\langle s\rangle}\simeq \beta_{m^2}\sim\frac{m^2_{soft}}{16\pi^2}.
\end{equation}
In the last expression we give an order of magnitude estimate: the
RG beta function for $m^2$ is the sum of several terms of order $m_{soft}^2$
(multiplied by some coupling constants), 
and part of the $16\pi^2$ suppression can be
compensated when all the terms are included.

In the case of stabilization by non-renormalizable terms,
\begin{equation}
M_S^2=2(K+1)(-m^2)\sim m^2_{soft}.
\end{equation}

In the preceding discussion, we have ignored the presence of scalar fields other
than $s_1$ and $s_2$ in the potential. In addition, there are extra degrees of
freedom
from the two singlets. The real field transverse to the flat direction  
eqn.~(\ref{flatreal}) is forced to take a very small VEV of order
$m^2_{soft}/\langle s\rangle$. The physical excitations along that transverse
direction have (up to soft mass corrections) an intermediate scale mass 
\begin{equation}
\label{intmass}
M_I^2={g_1'}^2(Q_1^2\langle s_1\rangle^2+
Q_2^2\langle s_2\rangle^2).
\end{equation}
The two pseudoscalar degrees
of freedom Im $S_1$, Im $S_2$ are massless: the potential is invariant under
independent rotations
of the phases of $S_1$ and $S_2$ so that the spontaneous breaking of this
$U(1)\times U(1)$ symmetry gives two Goldstone bosons. One of the $U(1)$'s is
identified with the gauged $U(1)'$ and the corresponding Goldstone boson is
eaten by the $Z'$, which has precisely the same intermediate mass given by
eqn.~(\ref{intmass}). The other massless pseudoscalar remains in the physical
spectrum and can acquire a mass if there are  terms in the potential that 
break the other $U(1)$ symmetry explicitly (e.g., in the presence of 
$AW_{\rm NR}$ terms.).
The fermionic part of the $Z'-S_1-S_2$ sector consists of three neutralinos
$(\tilde{B}',\tilde{S}_1,\tilde{S}_2)$. The combination
\begin{equation}
\tilde{S}=\cos\alpha_Q \tilde{S}_1 + \sin\alpha_Q \tilde{S}_2
\end{equation}
is light, with mass of order $m_{soft}$ if the minimum is fixed by
non-renormalizable terms. If the minimum is instead determined by the running
of $m^2$, $\tilde{S}$ is massless at tree-level but acquires a mass at one-loop 
of order $m_{soft}/(4\pi)$. The two other
neutralinos have masses $M_I\pm\frac{1}{2}M_1'$,
where $M_1'$ is the $U(1)'$ soft gaugino mass.

This pattern of masses can be easily understood; in the absence of supersymmetry
breaking, a nonzero VEV along the flat direction breaks the $U(1)'$ gauge symmetry
but leaves supersymmetry unbroken. Thus, the resulting spectrum is arranged in
supersymmetric multiplets: one massive vector multiplet (consisting of the $Z'$ 
gauge vector boson, one real scalar and one Dirac fermion) has mass $M_I$, and one
chiral multiplet (consisting of the complex scalar $S$ and its Weyl fermion
partner $\tilde{S}$) remains massless. The presence of soft supersymmetry breaking
terms modifies the picture slightly, lifting the mass degeneracy of the
components in a given multiplet by amounts proportional to the soft breaking.

The rest of the fields that may be present in the model can be classified into two
types; those that couple directly in the renormalizable superpotential to
$\hat{S}_{1,2}$ will acquire intermediate scale masses, and those which do not
can be kept light. In particular, all the usual MSSM fields should belong to the
latter class. The particle spectrum at the electroweak scale thus contains the
usual MSSM fields and one extra chiral multiplet $(S,\tilde{S})$ remnant of the
$U(1)'$ breaking along the flat direction\footnote{In the case of a single $S$
and no additional $U(1)'$ there is also one extra chiral multiplet at the
electroweak scale.}. The interactions among the light
multiplet $\hat{S}$ and MSSM fields are suppressed by powers of the intermediate
scale. At the renormalizable level, the only interaction between the MSSM fields
and the intermediate scale fields arises from the $U(1)'$ $D$-terms in the scalar
potential. The resulting effect after integrating out the fields which have heavy
intermediate scale masses \cite{drees} is a shift of the soft masses of MSSM
fields charged under the extra $U(1)'$:  \begin{equation} \delta
m_i^2=-Q_i\frac{m_1^2-m_2^2}{Q_1-Q_2}.  \end{equation} The $U(1)'$ $D$-term
contribution to the scalar quartic coupling of light fields charged under the
$U(1)'$ drops out after decoupling these intermediate scale particles. 

Non-renormalizable interactions between MSSM fields and the $S_{1,2}$ fields,
which can play an important role (e.g., for the generation of the $\mu$ parameter
and fermion masses) are discussed separately in Section IV.

Before closing this section, we remark in passing that a similar intermediate
scale breaking can occur in the $H_{1,2}$ sector of the theory, where, in the
absence of a fundamental $\mu$ parameter, the direction $H_1^0=H^0_2$ is also
flat. The condition $m_{H_1}^2+m_{H_2}^2>0$ on the Higgs soft masses would prevent
the formation of such a dangerous intermediate scale minimum. This is however not
a necessary condition; the breaking could well occur first along the $S$ flat
direction generating an effective $\mu$ parameter that can lift the $H_1^0=H_2^0$
flat direction. The determination of which breaking occurs first would require
an analysis of the effective potential in the early Universe.

\section{Radiative Breaking}

Both mechanisms for fixing an intermediate VEV (purely radiative or by
non-renormalizable
terms) depend on the scale $\mu_{RAD}$ at which some combination of 
squared soft
masses is driven to negative values in the infrared. In this section we present
several examples in which the breaking of the extra $U(1)'$ can take place
naturally at an intermediate scale and examine the range of the scale
$\mu_{ RAD}$.

For the sake of concreteness, 
we consider three models in which one or both of the
singlets couples to exotic superfields in the renormalizable
superpotential:  
\begin{itemize} 
\item Model (I): 
$\hat{S}_1$ couples to exotic $SU(3)$ triplets $\hat{D}_1$, $\hat{D}_2$ in the
superpotential
\begin{eqnarray}
\label{lsect1}
W=h\hat{D}_1\hat{D}_2\hat{S}_1.
\end{eqnarray}

\item Model (II): 
$\hat{S}_1$ couples to exotic $SU(3)$ triplets $\hat{D}_1$, $\hat{D}_2$ and
$\hat{S}_2$ to exotic
$SU(2)$ doublets $\hat{L}_1$, $\hat{L}_2$ in the superpotential
\begin{eqnarray}
\label{lsect2}
W=h_D\hat{D}_1\hat{D}_2\hat{S}_1+h_L\hat{L}_1\hat{L}_2\hat{S}_2.
\end{eqnarray}

\item Model (III):
$\hat{S}_1$ couples to $N_p$ identical pairs of MSSM singlets, charged under
$U(1)'$,
in the superpotential
\begin{eqnarray}
\label{lsect3}
W=h\sum_{i=1}^{N_p}\hat{S}_{ai}\hat{S}_{bi}\hat{S}_1.
\end{eqnarray}
\end{itemize}

We have analyzed the renormalization group equations (RGEs) of each model to
determine the range of $\mu_{ RAD}$ as a function of the values of the parameters
at the string scale. 
In principle, we could consider other models, 
such as a variation of Model (II) 
in which the
same singlet couples to the exotic triplets and doublets through
$W=h_D\hat{D}_1\hat{D}_2\hat{S}_1+h_L\hat{L}_1\hat{L}_2\hat{S}_1$, or a variation
of Model (III) in which the
singlet couples to additional singlets that are not a set of $N_p$ 
identical pairs through $W=\sum_{i,j}C_{i,j}\hat{S}_i\hat{S}_j\hat{S}_1$.  
For simplicity, we restrict
our consideration to these three models, 
because they can be analyzed analytically\footnote{Even simpler analytic examples, neglecting trilinear $A$-terms, gaugino masses, and the running of the 
Yukawas, are given in the Appendix of \cite{CL}.}. 

We assume gauge coupling unification at $\MS$,
such that
\begin{eqnarray}
\label{lgcbc}
g^0_3=g^0_2=g^0_1=g_1'^0=g_0,
\end{eqnarray}
which is approximately consistent with the observed gauge coupling
unification\footnote{We assume a GUT normalization for the Abelian gauge
couplings, such that $g_1=\sqrt{k}g_Y$, where $g_Y$ is the coupling
usually called $g'$ in the Standard Model and $k= \frac{5}{3}$. 
In general, string models considered could have $k\ne \frac{5}{3}$.}.  
At the one-loop level, the 
singlets in Model (III)  do not affect
the gauge coupling unification of the MSSM.
Model (II) is also consistent with gauge
coupling unification if the $D_i$, $L_i$ are approximately degenerate in
mass,
because they have the appropriate quantum numbers to fit into multiplets of $SU(5)$.  
However, the presence of the exotic triplets not part of an $SU(5)$ multiplet
violates the gauge coupling unification in Model (I).  
This problem can be resolved
if there are other exotics which do not couple to $S_1$ but contribute to the
running of the gauge couplings (e.g., additional $SU(2)$ doublets; {\it i.e.},
Model (I) is a limiting case of Model (II) as $h_L$ goes to zero). The additional
exotics will generally have electroweak scale masses so they will not precisely
cancel the effects of the triplets except by accident. However, Model (I) is
still useful to illustrate the basic ideas.

For the sake of simplicity, we assume that the boundary conditions for the Yukawa
couplings are given
by 
\begin{eqnarray}
\label{lyukbc}
h^0=g_0\sqrt{2},
\end{eqnarray}
as calculated in string models based on fermionic ($Z_2\times Z_2$) orbifold 
constructions at a special point in moduli 
space\footnote{An overall normalization
factor of $g_0\sqrt{2}$ at the string scale is required if the 
three-gauge-boson
coupling is to be $g_0$. In this class of string models, cubic couplings in a
superpotential can contain additional factors of 
$(\frac{1}{\sqrt{2}})^n$, with $n\in \{0,1,2,3\}$.
The power $n$ corresponds to the number of
Ising fermion oscillator excitations paired with $\sigma_{+}\sigma_{-}$
factors (i.e., sets of order/disorder operators) present
in the product of vertex operators associated with the 
multiplets in the superpotential term.}.
Thus, the analysis presented below relies on large Yukawa couplings to exotic
fields, which are a generic feature of a class of string models considered.
However, the specific choice of exotic couplings in 
(\ref{lsect1}), (\ref{lsect2}), and (\ref{lsect3}) 
is chosen for concreteness in
order to illustrate different symmetry breaking scenarios.

In the analysis, we assume unification of gaugino masses at $\MS$,
\begin{eqnarray}
M^{0}_{3}=M^{0}_{2}=M^{0}_{1}=M_1'^0=M_{1/2}
\label{univgau}
\end{eqnarray}
and universal\footnote{We do not consider nonuniversal soft mass-squared
parameters, because it is possible to explore the range of $\mu_{ RAD}$ without
this additional complication.} scalar soft mass-squared parameters,
\begin{eqnarray}
m^{0\,2}_{i}=m_{0}^{2}.
\label{univmass}
\end{eqnarray}
The first and third models have only one trilinear coupling, with initial
value $A^0$.  
We do consider the possibility of nonuniversal trilinear couplings
$A^0_D$, $A^0_L$ in the analysis of the second model.

The RGEs of the models (\ref{lsect1})-(\ref{lsect3}) are presented in a general 
form in the Appendix\footnote{The running of the
$U(1)'$ gauge coupling depends on the charge assignments of all of the fields
in the theory, and so is highly model dependent.  For simplicity, we assume
that the $U(1)'$ charge assignments are such that the evolution of $g'_1$
is identical to that of $g_1$.}. We have solved the RGEs 
in each case for a range of boundary conditions to determine the
range of $\mu_{ RAD}$. Each of the models considered has the
advantage that 
it is possible to obtain exact analytical solutions to the RGEs, which
yield insight into the nature of the dependence of the parameters on their
initial values. Exact solutions \cite{exact} are possible in these models because
the RGEs for the Yukawa couplings are decoupled.  In more complicated cases, e.g.,
if the same singlet couples to both triplets and doublets, no simple exact
solutions exist.   
It is also useful to consider simpler semi-analytic solutions to the
RGEs, in which the running of the gauge couplings and gaugino masses is
neglected in the solutions of the RGEs of the other parameters.  The exact and
semi-analytic solutions are presented in Appendix A.  
The results of the renormalization
group analysis are presented in Tables 
I-III for Models (I)-(III), respectively.  The evolution of the parameters of
Model (I) is shown in some representative graphs.  

{\bf Model (I)}: In Table I, we present the results of the analysis of Model (I).
We first choose the  $U(1)'$ charge assignment
$Q_1=-Q_2=-1$ for the singlets $\hat{S}_1$ and $\hat{S}_2$ and investigate the
nature of
$\mu_{ RAD}$ as a function of the initial values of the dimensionless ratios
$A^0/m_0$ and $M_{1/2}/m_0$. The scale dependence of the Yukawa
coupling\footnote{The evolution of the Yukawa coupling for large initial values
demonstrates the fixed point behavior, as discussed in the Appendix.} and
the trilinear coupling are shown in Figure 1. With this 
choice of $U(1)'$ charges and
$A^0=m_0$, the breaking scale is of the order $10^{10}$ GeV for values
of $M_{1/2}={\cal O}(m_0)$. However, radiative breaking (along the $D$-flat
direction) is not
achieved for small values of the initial gaugino masses, as is also shown in
Figure 2~(a).  
\begin{figure}[hbt]
\centerline{\hbox{
\psfig{figure=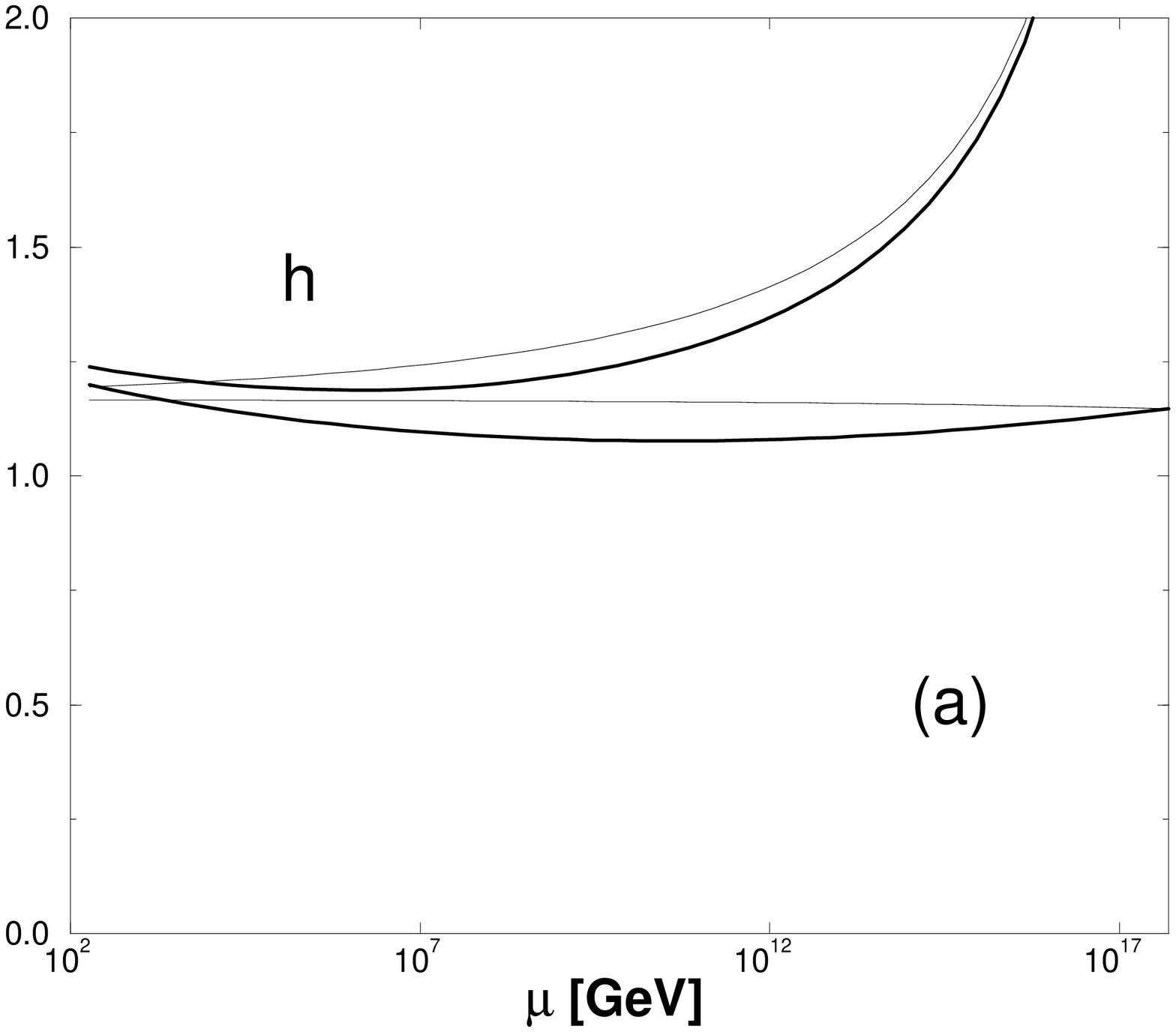,height=9cm,width=8cm,angle=-90,bbllx=3.cm,bblly=3.cm,bburx=21.cm,bbury=24.cm}
\psfig{figure=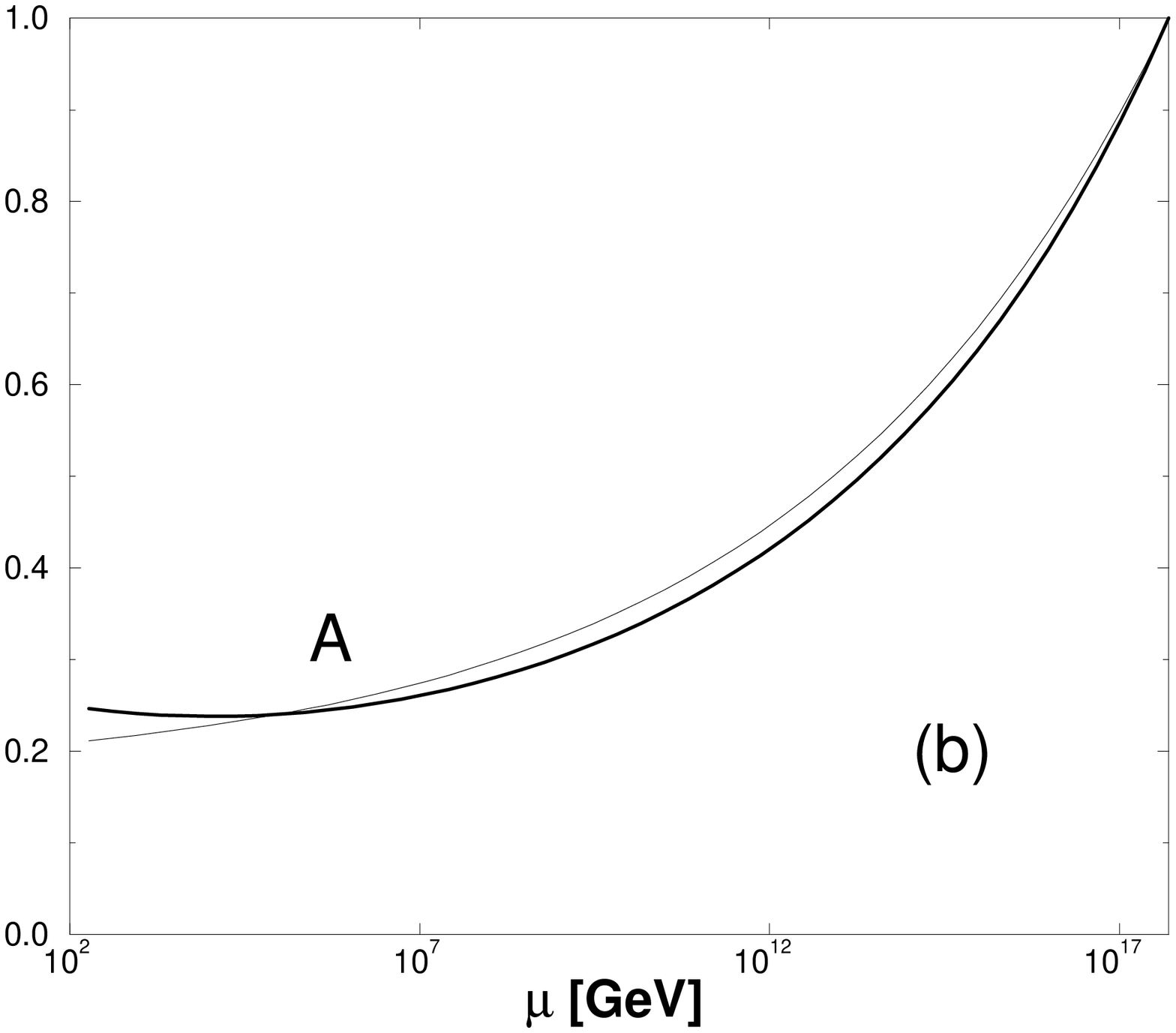,height=9cm,width=8cm,angle=-90,bbllx=3.cm,bblly=3.cm,bburx=21.cm,bbury=24.cm}}}
\caption{\footnotesize
(a) Scale dependence of the Yukawa coupling of Model (I) for $h^0=g_0\sqrt{2}$
and $h^0=10$.
(b) Scale dependence of the trilinear coupling of Model (I) in units of $m_0$,
with $M_{1/2}=0.1m_0$.  In each case $Q_1=-Q_2=-1$.  
Bold curves are for exact solutions, and light curves represent semi-analytic
approximations.}
\end{figure}
\begin{figure}[hbt]
\centerline{\hbox{
\psfig{figure=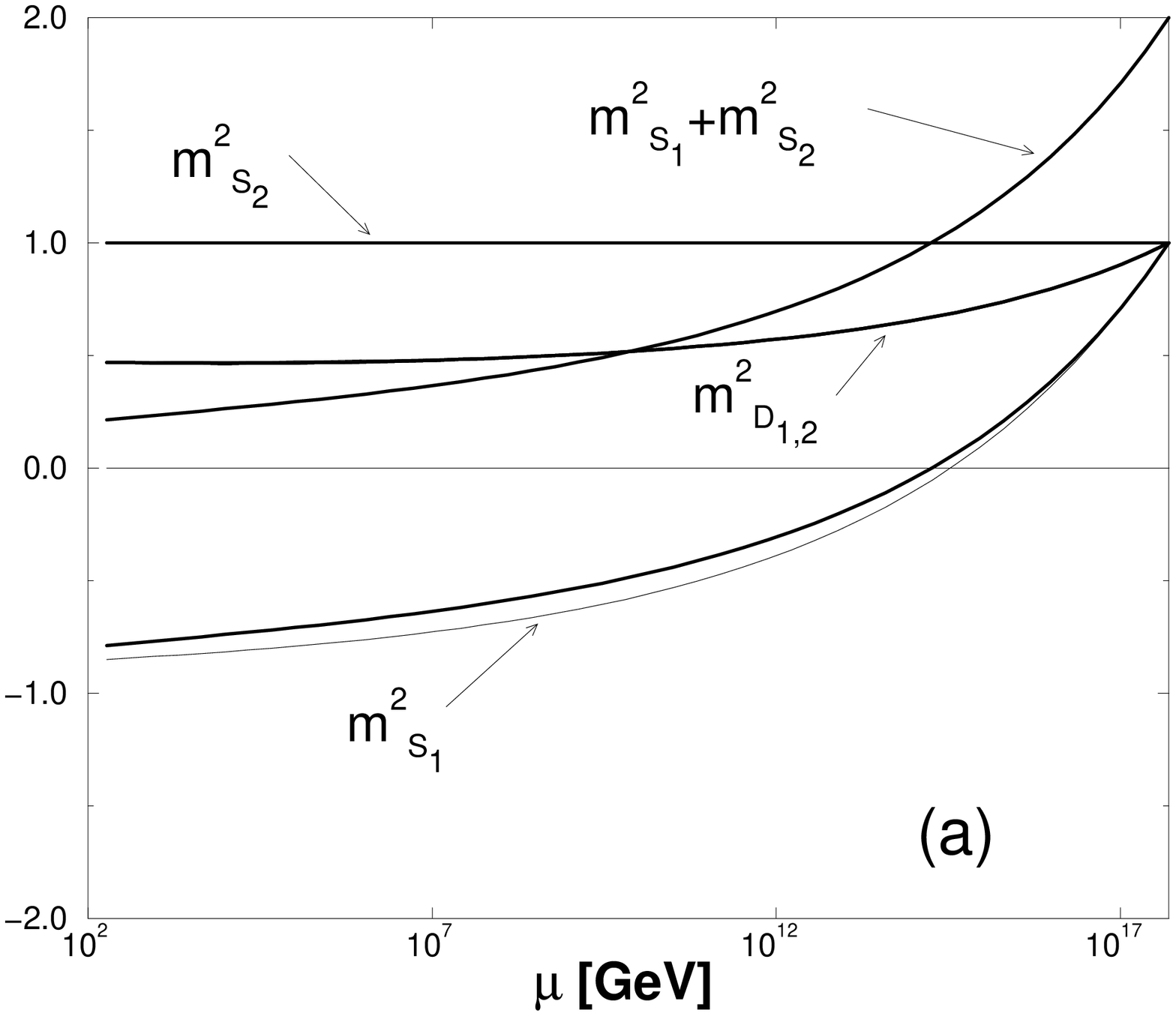,height=9cm,width=8cm,angle=-90,bbllx=3.cm,bblly=3.cm,bburx=21.cm,bbury=24.cm}
\psfig{figure=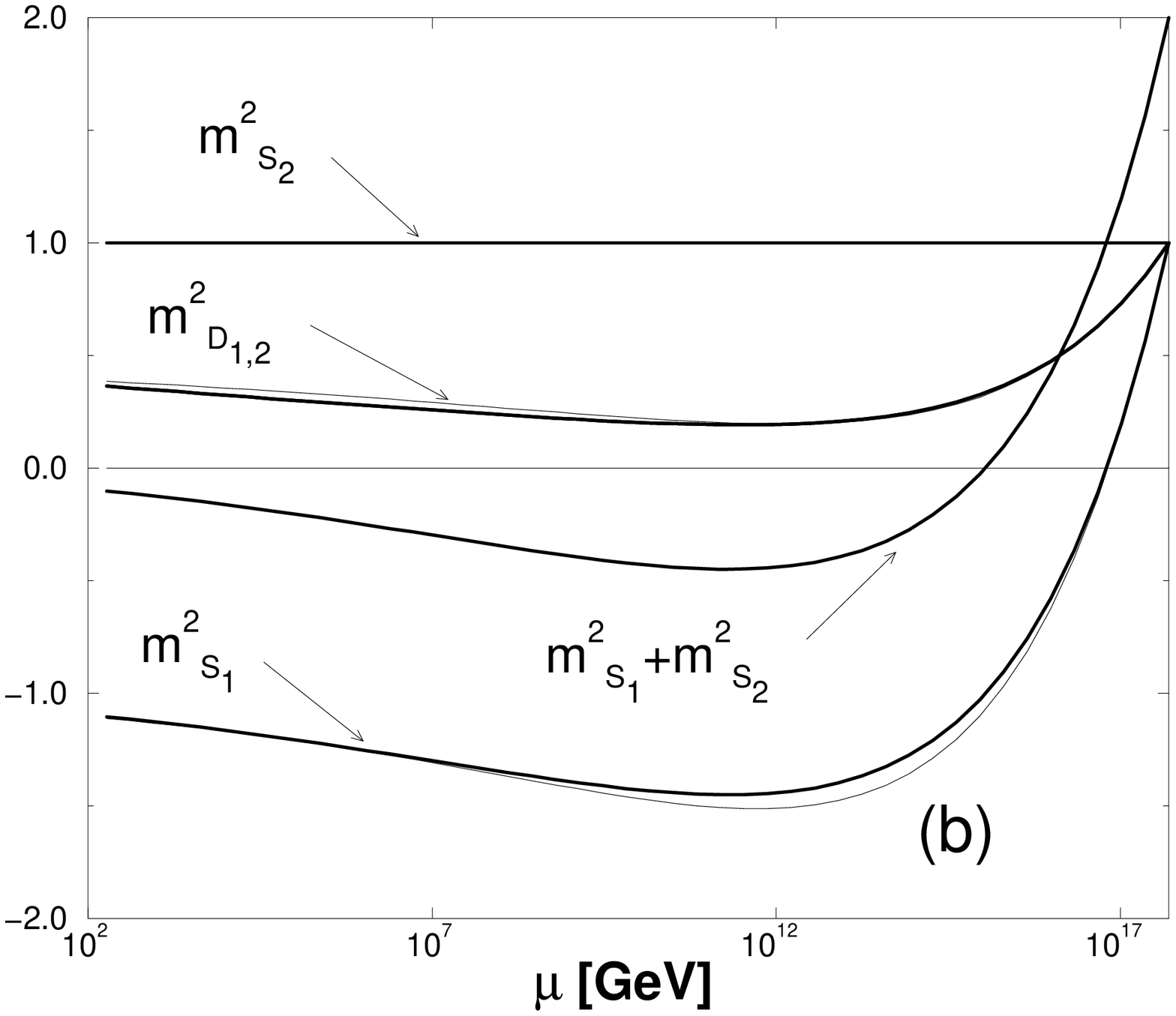,height=9cm,width=8cm,angle=-90,bbllx=3.cm,bblly=3.cm,bburx=21.cm,bbury=24.cm}}}
\caption{\footnotesize
(a) Scale dependence of the soft mass-squared parameters of Model (I) in units
of $m^2_0$, with
$A_0=m_0$ and $M_{1/2}=0.1m_0$. 
(b) same, except $A_0=3m_0$ and $M_{1/2}=0.1m_0$.  In each case $Q_1=-Q_2=-1$.  
Bold curves are for exact solutions, and light curves represent  semi-analytic
approximations.}
\end{figure}
The gaugino
mass parameter $M_{1/2}$ governs the fixed point behavior of the soft 
mass-squared parameters (as was also found in \cite{CDEEL}), such that small
gaugino masses do not drive $m_1^2$
sufficiently negative to overcome the fact that $m^2_2$ does not run
significantly because it does not have any couplings in the
superpotential. $S_1$ will acquire an electroweak scale VEV in this case, as was
described in Section~I. Increasing the value of $A^0$ increases $\mu_{ RAD}$
dramatically (up to $10^{17}$ GeV), for it drives $m^2_1$ negative at a
higher scale; this
behavior is also shown for the case of $A^0/m_0=3.0$, $M_{1/2}/m_0=0.1$ in
Figure 2~(b).  The breaking scale decreases significantly (in some cases, all
the way to the TeV range) when both $A^0/m_0$
and $M_{1/2}/m_0$ are lowered simultaneously.  This is to be expected, for
this is equivalent to raising the
initial value of the soft mass-squared parameters and keeping
$A^0=M_{1/2}$, in which case $m^2_1$ is driven negative at a lower scale.

For a given set of boundary conditions, it is also possible to raise or
lower $\mu_{ RAD}$ by choosing
different values of the ratio of $\mid\!Q_1/Q_2\!\mid$, as can be seen
from (\ref{msum}).  In particular, $\mid\!Q_1/Q_2\!\mid >1$ will increase the
relative weight of $m^2_2$ and so decrease $\mu_{ RAD}$, while     
$\mid\!Q_1/Q_2\!\mid<1$ will increase the relative weight of $m^2_1$
and thus increase $\mu_{ RAD}$.  Several examples of this type are
presented in Table I. The values of $\mu_{ RAD}$ for the examples with
$M_{1/2}/m_0=1.0$ should be contrasted with the value of $10^{10}$ GeV 
obtained with $Q_1=-Q_2=-1$, and the values with $M_{1/2}/m_0=0.1$ should 
be compared with the result that radiative breaking does not occur in
the case with equal and opposite $U(1)'$ charges. 

{\bf Model (II)}: The results of the analysis of Model (II) are presented in Table
II. In this
case, both $m^2_1$ and $m^2_2$ are driven negative due to the large Yukawa
couplings in the superpotential.  Thus, $\mu_{ RAD}$ is
generally much higher than in the case of the model previously discussed, 
of order $10^{13}$ to $10^{17}$ GeV for $Q_1=-Q_2=-1$.  The breaking scale
increases with larger values of $M_{1/2}$ and $A^0$, and  the
effects of the gaugino masses are negligible for sufficiently
large values of $A^0/m_0$.  The breaking scale can be lowered to
the range of $10^{11}$ GeV by decreasing the values of
$A^0/m_0$ and
$M_{1/2}/m_0$.  In this model, changing the value
of $\mid\!Q_1/Q_2\!\mid$ does not have a significant effect on
$\mu_{ RAD}$, as the soft mass-squared parameters of both singlets are
driven to negative values.

{\bf Model (III)}: In Table III, we present the results of the analysis of Model (III),
in which $S_1$ couples to identical pairs of singlets charged under the
$U(1)'$, and
$S_2$ has no couplings in the renormalizable superpotential.  In this
case, the number of pairs of singlets is analogous to the group theoretical weight
in the RGEs, such
that $m^2_1$ is driven negative at some scale.  While $N_p=3$ gives the
same weight as that of the first model with exotic triplets, the
values of $\mu_{ RAD}$ shown in the first two entries of Table III
demonstrate that this model does not mimic the first model.  
For example, the results show
that to obtain radiative breaking for $A^0/m_0=1.0$, it is
necessary to take large values of $N_p$ (such as $N_p=7$ for
$\mu_{ RAD}\sim10^4$ GeV). This is due to the fact that Model (III) does not
have the $SU(3)$ coupling, so all of the parameters have a smaller gauge
contribution.   In particular, the Yukawa coupling is weaker in Model (III),
so $m^2_1$ is not driven to negative values as quickly as in Model (I).  This
model also differs from the previous models in that 
smaller values of $M_{1/2}$ yield larger values of $\mu_{ RAD}$, often by
many orders of magnitude. Increasing
the value of $A^0$ raises the breaking scale dramatically even for small
values of $N_p$, eventually dominating the effects of the gaugino masses.

Several examples are also presented in Table III in which the breaking scale is
modified by choosing different values of $\mid\!Q_1/Q_2\!\mid$ for a given set
of boundary conditions. 
As in the first model, the scale can be raised significantly 
(e.g., to $10^{5}$ GeV from the case of no solution)
by assigning charges such that 
$\mid\!Q_1/Q_2\!\mid<1$. The value of $\mu_{ RAD}$ can also be lowered
substantially (e.g., to $10^{4}$ GeV from $10^{11}$ GeV) by choosing 
$\mid\!Q_1/Q_2\!\mid>1$.

The results of this analysis demonstrate that within models in which only one of
the singlets couples to exotic matter in the renormalizable superpotential (such
as Model (I) and Model (III)), there is a broad range of values of the breaking
scale $\mu_{ RAD}$, from the TeV range up to around $10^{16}$ GeV. In many cases
there is no $D$-flat solution, so that the $U(1)'$ breaking will be at the
electroweak scale. While the non-renormalizable terms will be important if
$\mu_{ RAD}$ is sufficiently high, in many cases the scale of radiative breaking
will determine the VEV of $S$.  If both singlets have trilinear couplings in the
superpotential (such as in Model (II)), the breaking is strongly radiative, such
that non-renormalizable terms will dominate the symmetry breaking.

\begin{table}[tbp]
\begin{center}
\begin{tabular}{||c|c|c|c||c|c|c|c||}
$Q_1$,$Q_2$&$A^0/m_0$ & $M_{1/2}/m_0$ & $\mu_{ RAD}$(GeV) &$Q_1$,$Q_2$&$A^0/m_0$
& $M_{1/2}/m_0$ & $\mu_{ RAD}$(GeV) \\ \hline
$-$1,1&1.0& 1.0 & $2.7\times10^{10}$&$-$1,1&0.3& 0.3 & $2.0\times10^{3}$ \\
$-$1,1& 1.0&0.1& -- &$-\frac{5}{4}$,$\frac{3}{4}$&1.0&1.0&$3.3\times10^8$\\
$-$1,1&3.0& 1.0&$1.5\times10^{15}$
&$-\frac{5}{4}$,$\frac{1}{3}$&1.0&1.0&$2.1\times10^4$\\
$-$1,1&3.0&0.1&$1.1\times10^{15}$&$-$1,$\frac{1}{3}$&1.0&1.0&$2.3\times10^5$\\
$-$1,1&5.0& 1.0&$8.8\times10^{16}$&$-$1,$\frac{1}{4}$&1.0&1.0&$2.1\times10^3$\\
$-$1,1&1.0& 0.5 & $2.6\times10^{7}$ &$-\frac{1}{2}$,1&1.0&0.1&$6.0\times10^8$\\
$-$1,1&0.5& 0.5 & $2.4\times10^{6}$ &$-\frac{1}{3}$,1&1.0&0.1&$1.5\times10^{11}$\\
$-$1,1&0.7& 0.7 &
$3.2\times10^{8}$&$-\frac{1}{3}$,$\frac{1}{2}$&1.0&0.1&$2.3\times10^{5}$
\end{tabular}
\end{center}
\caption{ \footnotesize Model (I). Singlet coupled to triplets:
$W=h\hat{D}_1\hat{D}_2\hat{S}_1$.}
\end{table}

\begin{table}[tbp]
\begin{center}   
\begin{tabular}{||c|c|c|c||c|c|c|c||}
$Q_1$,$Q_2$&$A^0_D/m_0$,$A^0_L/m_0$ & $M_{1/2}/m_0$ &
$\mu_{ RAD}$(GeV)&$Q_1$,$Q_2$&$A^0_D/m_0$,$A^0_L/m_0$ & $M_{1/2}/m_0$ &
$\mu_{RAD}$(GeV) \\ \hline
$-$1,1&1.0,1.0& 1.0 & $1.0\times10^{14}$&$-$1,1&1.0,3.0&0.1&$4.4\times10^{15}$\\
$-$1,1&1.0,1.0& 0.1& $1.1\times10^{13}$&$-$1,1&3.0,3.0& 1.0&$4.0\times10^{16}$\\
$-$1,1&3.0,1.0& 1.0&$1.1\times10^{16}$&$-$1,1&5.0,5.0& 1.0&$1.9\times10^{17}$\\
$-$1,1&3.0,1.0&0.1&$1.1\times10^{16}$&$-$1,1&0.3,0.3& 0.3&$1.9\times10^{12}$\\
$-$1,1&1.0,3.0& 1.0&$4.4\times10^{15}$&$-$1,1&0.1,0.1& 0.1&$7.4\times10^{11}$
\end{tabular}    
\end{center}
\caption{ \footnotesize Model (II). Singlets coupled to triplets and doublets:
$W=h_D\hat{D}_1\hat{D}_2\hat{S}_1+h_L\hat{L}_1\hat{L}_2\hat{S}_2$.}
\end{table}

\begin{table}[tbp]
\begin{center}
\begin{tabular}{||c|c|c|c||c|c|c|c||}
$Q_1$,$Q_2$,$N_{p}$&$A^0/m_0$ & $M_{1/2}/m_0$ &
$\mu_{RAD}$(GeV)&$Q_1$,$Q_2$,$N_{p}$&$A^0/m_0$ & $M_{1/2}/m_0$ & $\mu_{RAD}$(GeV)
\\ \hline
$-$1,1,3&1.0& 1.0 &-- &$-$1,1,8&1.5&0.1&$3.9\times10^{11}$\\
$-$1,1,3&1.0& 0.1 &--&$-$1,1,10&1.5&1.0&$8.7\times10^{11}$\\
$-$1,1,7&1.0& 1.0 &--&$-$1,1,3&3.0& 1.0& $7.9\times10^{12}$ \\
$-$1,1,7&1.0& 0.1 & $3.4\times10^4$&$-$1,1,3&3.0& 0.1&$3.8\times10^{13}$ \\
$-$1,1,8&1.0& 1.0 &  --&$-$1,1,4&3.0&1.0&$4.8\times10^{14}$\\
$-$1,1,8&1.0& 0.1 & $3.5\times10^7$&$-$1,1,4&3.0& 0.1&$4.2\times10^{15}$\\
$-$1,1,10&1.0& 1.0 & $1.4\times10^7$&$-\frac{1}{3}$,1,3&1.0&0.1&$6.4\times10^4$\\
$-$1,1,10&1.0& 0.1
&$1.1\times10^{10}$&$-\frac{1}{2}$,1,3&1.5&0.1&$1.0\times10^5$\\
$-$1,1,3&1.5&1.0& --&$-1,\frac{4}{5}$,8&1.5& 0.1&$4.4\times10^{4}$\\
$-$1,1,8&1.5&1.0& $3.2\times10^8$&$-1,\frac{2}{3}$,4&3.0& 0.1&$3.6\times10^{12}$\\
\end{tabular}
\end{center}
\caption{ \footnotesize Model (III). Singlet coupled to singlet pairs:
$W=h\sum_{i=1}^{N_p}\hat{S}_{ai}\hat{S}_{bi}\hat{S}_1$.}
\end{table}

\section{Intermediate Scale due to Non-Renormalizable
Terms in  String Models}

\label{flatsection}
The scenarios discussed in  a general  particle physics
context  in the previous sections have
interesting implications for string models.
In particular, in a large class of string models the particle spectrum
consists of  SM singlets  $S_i$  whose (particular combination) ensures that they
correspond to $D$-flat directions and $F$-flat directions at least for the
renormalizable terms in the superpotential. On the other hand, it is often the
case  that these fields do have  non-renormalizable terms in the
superpotential, which along with the radiatively induced negative 
mass-square terms yield
intermediate scales with
implications for the SM sector of the theory.

For specific examples we shall concentrate on the type of
non-renormalizable terms in a class of  fermionic constructions. 
In such models, there are a number 
of SM singlets $S_i$ which are in general charged under additional 
$U(1)'$ factors.  The $D$-flatness  is ensured if  the $U(1)'$  charges of at
least two  $S_i$'s have   opposite signs. For the sake of concreteness we
confine ourselves
to the case of two $S_i$'s, with the $D$-flatness constraint  satisfying
eqn.~(\ref{flatdir}).
Since  the $S_i$  are massless states at $\MS$ they have 
no bilinear terms in the  superpotential. 
We  also require that in the superpotential
the trilinear self-couplings of  $S_i$  and the trilinear 
terms of  one $ S_i$   to  the  MSSM particle content 
are absent  as well.  This is often the case due to  
either (world-sheet) selection rules (as demonstrated below)  
and/or target space gauge coupling unification.

The analysis of the previous section has shown that  couplings to
exotic particles  with Yukawa coupling of ${\cal O}(1)$ can ensure a
radiative breaking for $S_i$'s. 
On the other hand, in general there are  
 non-renormalizable self-couplings 
 of $\hat S_i$'s in the superpotential.
It is convenient to rewrite (\ref{nrsup}) 
as\footnote{In these terms we have already 
chosen the $D$-flatness constraint,
and thus the non-renormalizable self-coupling is expressed in terms of the 
$S$ field (defined in (\ref{flatdir})) only.}
\begin{equation}
W_{K}= \hat{S}^{3} \left({\hat{S}\over \MK}\right)^K,
\label{nonren}
\end{equation}
where we have absorbed the coefficient $\alpha_K$ in the 
definition of mass scale $M$.
($M$ is related to $\cal M$ in eq.~(\ref{potflat})
as ${\cal M}= {\cal C}_K M$.) 
For simplicity we have not displayed the dependence of $M$ on $K$
or the detailed form of the operators.

If $p_1$ and $p_2$ are the unique relative primes defined by
$\frac{p_2}{p_1}=\frac{|Q_1|}{|Q_2|}$,
then, as discussed in section~II, $U(1)'$ invariance permits
values of $K>0$ in (\ref{nonren}) such that $3+K= (p_1+p_2)n$, where $n$ is
integer. World-sheet selection rules further constrain $K$ through 
restrictions on $n$ \cite{cvetic87,FINQ,rizos91,kalara,kobayashi}. 
For example, in the free fermionic
construction $n$ must be an even integer in the case of only two $S_i$, 
thus limiting $K$ to only odd values, independent of the values of the
$p_i$.

Fermionic world-sheet selection rules further require that both
singlets $S_i$ must originate from twisted 
world-sheet supersymmetric ({\it i.e.,}~Ramond) sectors of a model
for {\it any}  non-renormalizable   
terms of the form (\ref{nonren}) to appear in the superpotential.
In contrast, for a renormalizable trilinear
self-coupling $\hat S^3$ term to appear, one
of the two $S_i$ must have its origin in the untwisted Neveu-Schwarz
sector while the other comes from a Ramond sector \cite{rizos91,kalara}. 
Thus, renormalizable $(K=0)$ and non-renormalizable $(K>0)$  
terms of the form (\ref{nonren}) are mutually exclusive. 

The coefficients of the non-renormalizable couplings can be calculated in 
a large class of string models.  For the free fermionic construction, 
coefficient values can be cast in terms of the $K+3$--point string 
amplitude, $A_{K+3}$, in the the following 
form:\footnote{For the explicit calculation of the non-renormalizable terms in
a class of fermionic models, see \cite{kalara,faraggi96a}.}
\begin{eqnarray}
\left({1\over \MK}\right)^K
& \equiv &  \left( {\frac{\alpha_K}{M_{Pl}}}\right)^{K},\label{masnon4}\\
& = & {\left(2\alpha'\right)}^{K/2} A_{K+3}             \label{masnon1}\\
          & = &  {\left(2\alpha'\right)}^{K/2} 
                 \left(\frac{g}{2\pi}\right)^{K} 
                 g\eta C_K I_K
\label{masnon2}\\
          & = &  M_{Pl}^{-K} \left( \frac{4}{\sqrt{\pi}}\right)^{K}
                 g\eta C_K I_K
\label{masnon3}\\
\end{eqnarray}
where $g$ is the gauge coupling at $\MS$,
$\eta= \sqrt{2}$ is a normalization factor 
(defined so that the three-gauge-boson  
and two-fermion--one-gauge-boson couplings are simply $g$),
$2\alpha'\equiv (64\pi)/(M_{Pl}^2g^2)$  is the string
tension \cite{K}, $C_K$ is the coefficient of
${\cal O} (1)$ that encompasses different renormalization factors in the 
operator product expansion (OPE) of the string vertex operators (including the
target space gauge group Clebsch-Gordon coefficients), and 
$\alpha_K \equiv (4/\sqrt{\pi})^{K} g\eta C_K I_K$.
$I_K$ is the
world-sheet integral of the type:
\begin{equation}
I_K=\int d^2z_3\cdots d^2z_{K+2} 
f_K(z_1=\infty,z_2=1,z_3,\cdots,z_{K+2},z_{K+3}=0),
\label{igen}
\end{equation}
where $z_i$ is the world-sheet coordinate of the vertex operator of the
$i^{\rm th}$ string state. As a function of the
world-sheet coordinates, $f_K$ is 
a product of correlation functions formed respectively from the 
spacetime kinematics, Lorentz symmetry, ghost charge, 
local non-Abelian symmetries, local and global $U(1)$ symmetries, and
(non)-chiral Ising model factors in each of the vertex operators for
the $3+K$ fields.
All correlators but the Lorentz and 
Ising ones are of exponential form. For
non-Abelian symmetries and for $U(1)$ symmetries and ghost systems
these exponential correlators have the respective generic forms 
\begin{equation}
\langle
      \prod_{i} {\rm e}^{i \vec Q_i\cdot \vec J}
\rangle
=     \prod_{i<j} z_{ij}^{\vec Q_i \cdot \vec Q_j}  
\quad\quad {\rm and} \quad\quad 
\langle
      \prod_{i} {\rm e}^{i Q_i H}
\rangle
=     \prod_{i<j} z_{ij}^{Q_i Q_j}  
\label{expcor}
\end{equation}
where $z_{ij} = z_i - z_j$. In this language, $Q_i$ is imaginary for ghost 
systems.

While the Lorentz correlator is non-exponential, it
is nevertheless trivial and contributes a simple factor of
$z_{12}^{-1/2}$ to $f_K$. On the other hand,
the various Ising correlators are generically non-trivial.  
This makes $I_K$ difficult to compute.
In fact, Ising correlators generally
prevent a closed form expression for an integral $I_K$
\cite{kalara}.
Nevertheless, Ising fermions may be 
necessary in fermionic models for obtaining realistic gauge groups 
and (quasi)-realistic phenomenology \cite{kalara,CHL}.
Thus, although the Ising correlation functions 
make $I_K$ increasingly difficult to compute
as $K$ grows in value, Ising correlation functions
generally enter string amplitudes.
 
From \cite{kalara,faraggi96a} we infer that $I_1\sim 70$ 
and 
$I_2\sim 400$. In \cite{kalara,faraggi96a},
the non-renormalizable terms 
for which $I_1$ and $I_2$ were calculated 
involved only one and two MSSM singlets, respectively. 
However, we do not expect that the values of
$\IS_1$ or $\IS_2$ associated with terms composed 
totally of $S$-type singlets will generically vary 
significantly 
from the values obtained when some non-singlets are involved.

For a $K=1$ term composed solely of non-Abelian 
singlets\footnote{The four states forming this 
$K=1$ superpotential term were denoted 
$H_{30}$, $H_{32}$, $H_{37}$, and $H_{39}$ in Table 2 
of \cite{faraggi90a}. 
The first two of these states originate in one sector of the model, while
the latter two reside in a second sector. This is the general pattern 
also followed in \cite{kalara,faraggi96a}.}
carrying $U(1)'$ charge \cite{faraggi90a},
we have explicitly calculated a value for $\IS_1$.
For comparative purposes we relate our $\IS_1$ to the 
associated four-point string amplitude $\AS_4$ 
via the normalization,
\begin{equation}
\AS_4 = \frac{g}{2\pi}\frac{1}{4}\IS_1.
\label{eqncomp}
\end{equation}
This is the same normalization as in \cite{faraggi96a},
where the value of $I_1$ was $77.7$. The four singlet case produces  
\begin{equation}
\IS_1 =  2\sqrt{2}\int {d}^2 z\,\, | z |^{-1} | 1 - z |^{-3/2} . 
\label{i4usaz}
\end{equation}
By shifting $z\rightarrow z + 1$ and 
converting the world-sheet coordinate $z$
to polar coordinates $(r,\theta)$, the integral can be expressed as
\begin{equation} 
\IS_1  = 4\sqrt{2}\int^{\infty}_{0} dr\,\int^{\pi}_{0} d\theta \,\, 
\frac{1}{1+r^2 -2r\cos\,\theta }.
\label{i4usaro}
\end{equation}
Integrating over the angle $\theta$ results in     
\beqn
\IS_1  &=& 4\sqrt{2}\int^{\infty}_{0} dr\,\, 
\frac{1}{r}\frac{2\sqrt{r}}{r+1} K\left( \frac{2\sqrt{r}}{r+1}\right) 
\label{i4usar}\\
       &=& 8\sqrt{2}\int^{\infty}_{0} dl\,\, 
\frac{2}{l^2+1} K\left( \frac{2l}{l^2+1}\right) ,
\label{i4usarb}
\eeqn
where $K$ is the complete elliptic integral of the first 
kind.

Numerical approximation of (\ref{i4usaro}) 
(after splitting integration over $r$ into two separate regions 
$0\leq r \leq 1$ and $0\leq r \leq \infty$) 
via Mathematica yields a value of $\IS= 63.7$.
As a test of the numerical approximation, we can also 
expand $K$ in powers of $\frac{2\sqrt{r}}{r+1}$ 
(or in powers of $\frac{2 l}{l^2+1}$) and then 
integrate the first two (or more) terms in this series.
This latter approach yields (for two terms)  
$\IS_1\approx  (9/\sqrt{2})\pi^2 \approx 62.8 \pm 10\%$, 
in excellent agreement with the our numerical 
approximation.\footnote{$x\equiv {2\sqrt{r}}/{(r+1)}$ 
is within the range of convergence $0\le x < 1$
of the series expansion, 
\beq
K(x)= \frac{\pi}{2}
\left\{ 
1 + (\frac{1}{2}) x  + (\frac{1\cdot 3}{2\cdot 4})^2 x^2 + \cdots 
\right\}\, , 
\label{kx1}
\eeq 
for all values of $r$ except for $r=1$. At $r=1$, $x$ reaches the endpoint of 
convergence, $x=1$, for which $\lim_{x\rightarrow 1} K(x) \rightarrow \infty$. 
As consistency between our two estimates of  $\IS_1$ indicates, 
inclusion of this endpoint in the range of 
integration of the series expansion still permits using the series expansion.} 
Thus the non-singlet factor in the four-point string amplitude of  
\cite{faraggi96a} causes $I_1$ to be about $20\%$ 
larger than $\IS_1$. 

It is expected that the interference terms in $I_K$ are 
generically such that
$I_K < I_1^K$, and thus  ${\MK}>M_1$.
In particular, for  $K=1$ we obtain: $M_1\sim 3\times 10^{17}$ GeV 
using $I_1\sim 70$ 
and for $K=2$: $M_2\sim 7\times 10^{17}$ GeV using 
$I_2\sim 400$.

\section{NON-RENORMALIZABLE COUPLING TO THE MSSM PARTICLES}

The flat direction $S$ can have a set of non-renormalizable couplings
to MSSM states that offer solutions to the $\mu$ problem \cite{kini}
and yield mass hierarchies between generations \cite{hier}. 
The non-renormalizable $\mu$-generating terms are of the form,
\begin{equation}
\label{supermu}
W_{\mu}\sim \hat{H_1} \hat{H_2} \hat{S} \left( {\hat{S}\over \MK} \right)^P.
\label{nonrenmu}
\end{equation}
In addition, the effective soft SUSY-breaking $B$-term,  
$BH_1H_2+{\mathrm H.c.}$ in the Higgs potential,
which is necessary for a correct electroweak symmetry breaking, can appear
via mixed $F$-terms from a superpotential\footnote{Although 
the values of the $\MK$ in the two terms of eqn.~(\ref{nonrenmu2})
are expected to be of the same order of magnitude, they may vary 
somewhat. For simplicity, we ignore the distinction.}
\begin{equation}
W_B\sim \hat{H}_1 \hat{H}_2 \hat{S} \left( {\hat{S}\over \MK}\right)^P 
          +     \hat{S}^{3}       \left( {\hat{S}\over \MK}\right)^K ,
\label{nonrenmu2}
\end{equation}
or from supersymmetry breaking terms \cite{recmu} in the potential of the type
\begin{equation}
V\sim A H_1 H_2 S \left( {S\over \MK} \right)^P + {\mathrm H.c.}, 
\label{potmu1}
\end{equation}
where $A\sim m_{soft}$. In both cases, when the effective $\mu$ parameter
is of the order of the electroweak scale, $B\sim m^2_{soft}$ automatically.

Generational up, down, and electron mass terms appear,
respectively, via
\begin{equation}
W_{u_i}\sim \hat{H}_2 \hat{Q}_i \hat{U}^c_i 
            \left( {\hat{S}\over \MK} \right)^{P^{'}_{u_i}};\quad
W_{d_i}\sim \hat{H}_1 \hat{Q}_i \hat{D}^c_i 
            \left( {\hat{S}\over \MK} \right)^{P^{'}_{d_i}};\quad
W_{e_i}\sim \hat{H}_1 \hat{L}_i \hat{E}^c_i 
            \left( {\hat{S}\over \MK} \right)^{P^{'}_{e_i}},\quad
\label{nonrenql}
\end{equation}
with $i$ denoting generation number\footnote{Alternatively,
non-renormalizable chiral supermultiplet mass terms can be generated through
anomalous $U(1)'$ breaking \cite{F,binetruy96}. Typically, in that case the 
analogue
of $\langle s\rangle/M\sim 1/10$, so that larger values of $P'$ are required.}.

Majorana and Dirac neutrino terms may also be present via,
\begin{eqnarray}
&W&^{\rm (Maj)}_{L_iL_i}\sim 
              \frac{\left(\hat{H}_2 \hat{L}_i \right)^2}{\MK}
              \left( {\hat{S}\over \MK} \right)^{P^{''}_{L_iL_i}};\quad
W^{\rm (Dir)}_{L_i{\nu}^c_i}\sim \hat{H}_2 \hat{L}_i \hat{\nu}^c_i 
            \left( {\hat{S}\over \MK} \right)^{P^{'}_{L_i {\nu}^c_i}};
\nonumber\\
&W&^{{\rm (Maj)}}_{{\nu}^c_i{\nu}^c_i}\sim \hat{\nu}^c_i \hat{\nu}^c_i \hat{S}
\left( {\hat{S}\over \MK} \right)^{\bar{P}_{{\nu}^c_i{\nu}^c_i}}.\quad
\label{nonrenqlb}
\end{eqnarray}
($\hat{\nu}\in \hat{L}$ represents the neutrino doublet component and
we have introduced neutrino singlets $\hat{\nu}^c$.) 

When the VEV $\langle S\rangle$ is fixed solely by the running of $m^2$,
the size of the $\mu$ parameter will be determined by the scale $\mu_{RAD}$
and the value of $P$ in eqn.~(\ref{supermu}),
$\mu_{eff}\sim\frac{\mu_{RAD}^{P+1}}{M^P}$. 
For example, for $P=1$ a reasonable ${\mu}_{eff}\sim 1$ TeV would 
correspond to 
$\mu_{RAD}\sim 10^{10}$ GeV.
On the other hand,
concrete order of magnitude estimates can be made 
when the VEV is fixed by non-renormalizable self-interactions of $S$.
Generally, if $\mu_{RAD}\ll 10^{12}$ GeV running is the dominant
factor; whereas, 
if $\mu_{RAD}\gg 10^{12}$ GeV the non-renormalizable operators
(NRO) dominate instead. 
With NRO-dominated $\langle S\rangle \sim (m_{soft} {\MK}^K)^{\frac{1}{K+1}}$,
the effective Higgs $\mu$-term takes the form,
\begin{equation}
 \mu_{eff} \sim m_{soft} \left( {m_{soft}\over \MK} \right)^{\frac{P-K}{K+1}}.
\label{nonrenmu3}
\end{equation}
The phenomenologically preferred choice among such  
terms is clearly $P=K$, yielding a $K-$independent 
${\mu}_{eff} \sim m_{soft}$.  Both of these intermediate scale scenarios
are to be contrasted to the case in which $\langle S \rangle$ is at the 
electroweak
scale. Then, $\mu_{eff}\sim m_{soft}$ can be generated by a renormalizable
$(P=0)$ term\cite{CDEEL}.

Quark and lepton masses can have hierarchical patterns 
generated through
\begin{equation}
m_{u_i} \sim \langle H_2\rangle
\left(\frac{m_{soft}}{\MK}\right)^{\frac{P^{'}_{u_i}}{K+1}};
\quad  
m_{d_i} \sim \langle H_1\rangle
\left(\frac{m_{soft}}{\MK}\right)^{\frac{P^{'}_{d_i}}{K+1}};  
\quad
m_{e_i} \sim \langle H_1\rangle
\left(\frac{m_{soft}}{\MK}\right)^{\frac{P^{'}_{e_i}}{K+1}}.  
\label{nonrenqlc}
\end{equation}
In (\ref{nonrenqlc}) we ignore the running of the effective Yukawas
below $\langle S \rangle$ (or below $\MP$ for $m_t$) because such effects 
are small compared to the uncertainties in $M$.

Comparison of the physical fermion mass ratios \cite{barnett96}
in Table \ref{masstable1} with 
theoretical $K$ and $P$ dependent mass values in
Table \ref{masstable2} suggests that the set 
\begin{eqnarray}
P^{'}_{1}&\equiv& P^{'}_{u_1}=P^{'}_{d_1}=P^{'}_{e_1} = 2;
\nonumber\\
P^{'}_{2}&\equiv& P^{'}_{u_2}=P^{'}_{d_2}=P^{'}_{e_2} = 1; 
\label{psets}
\end{eqnarray}
when used in tandem with $K=5$ or $K=6$,
could produce a fairly realistic 
hierarchy for the first two generations in the 
${\rm tan} \beta \equiv \frac{\langle H_2\rangle}{\langle H_1\rangle}\sim 1$ 
limit\footnote{In Table \ref{masstable2} we have used the computed value of
$M_1\sim 3\times 10^{17}$ GeV as the value for all $M$. 
To test the validity of this approximation,
we have also determined 
$\frac{m_{Q,L}}{\langle H_i\rangle}$ and
$\frac{\mu_{eff}}{m_{soft}}$ 
for $K=2$ 
using $M_2\sim 7\times 10^{17}$ GeV
and for $K=3$ using an extrapolated $M_3$ value of
$11\times 10^{17}$ GeV. 
For $P<5$ and $P'<K+5$, 
the better estimates of $M_2$ and $M_3$  
reduce
$\frac{m_{Q,L}}{\langle H_i\rangle} $ and 
$\frac{\mu_{eff}}{m_{soft}}$, respectively, 
only by factors of ${\cal O}(1)$ in comparison to the values
of $\frac{m_{Q,L}}{\langle H_i\rangle} $ 
($\frac{\mu_{eff}}{m_{soft}}$) 
given in the $K=2$- and $K=3$-columns of  
Table \ref{masstable2}.
However, larger values of $P$ and $P'$
yield increasing significant reductions in
$\frac{m_{Q,L}}{\langle H_i\rangle}$ and 
$\frac{\mu_{eff}}{m_{soft}}$, respectively, 
when the better estimates of $M_2$ and $M_3$ are used.}.
Alternatively, taking the
${\rm tan} \beta \sim 50 $ limit would suggest 
slightly higher values for $K$ (while keeping the same set of $P'$ values). 

Presumably $m_t$ is associated with a renormalizable coupling 
($P'_{u_3}= 0$). The other third family masses do not fit quite as well: 
they are 
too small to be associated with renormalizable couplings, but somewhat larger 
than is expected for $P'_{d_3}= P'_{e_3}= 1$ for $K=5$ or $K=6$. However, 
given the roughness of the estimates and the simplicity of the model, the 
overall pattern of the 
masses is quite encouraging. It is also possible that $m_b$ and $m_{\tau}$ are
associated with some other mechanism, such as non-renormalizable operators 
involving the VEV of an entirely different singlet.

There is an obvious constraint on a string model that could   
produce a generational mass hierarchy along these lines, containing
$P'_1 - 1 = P'_2 = P'_{u_3} + 1 = 1$ fermion mass terms, 
in tandem with a $P = K = 5$ or $6$ $\mu$-term.
A combination of world-sheet selection rules and 
$U(1)'$ charges must prevent $\mu$-generating terms with $P<5$ from appearing, 
while allowing the low order
$P'_i$ fermion mass terms.  If $U(1)'$ charges could be assigned by fiat
to each state, then the $U(1)'$ symmetry should be able to accomplish this 
by itself.
However, $U(1)'$ charge assignments are related to modular invariance
and thus they cannot be freely chosen for many states. 
World-sheet selection rules must likely play a role in constraining $P$.  


\begin{table}
\begin{tabular}{ccccccccccc}
$m_u$
&$:$
&$m_c$
&$:$
&$m_t$
&$=$ 
&$3\times 10^{-5}$
&$:$
&$7\times 10^{-3}$
&$:$
&$1$
\\
\hline
$m_d$
&$:$
&$m_s$
&$:$
&$m_b$ 
&$=$
&$6\times 10^{-5}$
&$:$
&$1\times 10^{-3}$
&$:$
&$3\times 10^{-2}$
\\
\hline
$m_e$
&$:$
&$m_{\mu}$
&$:$
&$m_{\tau}$ 
&$=$
&$0.3\times 10^{-5}$
&$:$
&$0.6\times 10^{-3}$
&$:$
&$1\times 10^{-2}$
\\
\hline
$m_{\nu_e}$
&$:$
&$m_{\nu_\mu}$
&$:$
&$m_{\nu_\tau}$ 
&$=$
&$<6\times 10^{-11}$
&$:$
&$<1\times 10^{-6}$
&$:$
&$<1\times 10^{-4}$
\\
\end{tabular}
\caption{Fermion mass ratios with the top quark mass normalized to $1$. 
The values of $u-$, $d-$, and $s$-quark masses used in the ratios
(with the $t$-quark mass normalized to $1$ from an assumed mass of $170$ GeV) 
are estimates
of the $\overline{\rm MS}$ scheme current-quark masses at a scale 
$\mu\approx 1$ GeV. 
The $c$- and $b$-quark masses are pole masses. 
An additional mass constraint for stable light neutrinos
is $\sum_i m_{\nu_i} \le 6\times 10^{-11}$ (i.e., $10$ eV), based on  
the neutrino contributions to the mass density of the universe
and the growth of structure [34].}   
\label{masstable1}
\end{table}

\begin{table}
\begin{tabular}{c|c|c|c|c|c|c|c|c}
       ~& $P$ or $P'$ & $K=1$ & $K=2$ & $K=3$ & $K=4$ & $K=5$
                                      & $K=6$ & $K=7$\\
\hline
\hline
 
$\left(\frac{m_{soft}}{\MK}\right)^{\frac{1}{K+1}}$
&
& $2\times 10^{-8}$
& $7\times 10^{-6}$
& $1\times 10^{-4}$
& $8\times 10^{-4}$
& $3\times 10^{-3}$
& $6\times 10^{-3}$
& $1\times 10^{-2}$
      \\
\hline
$\langle S\rangle$ (GeV)
& 
& $5\times 10^{9}$
& $2\times 10^{12}$
& $4\times 10^{13}$
& $2\times 10^{14}$
& $8\times 10^{14}$
& $2\times 10^{15}$
& $3\times 10^{15}$
      \\
\hline
& $K-1$
& $5\times 10^{7}$
& $1\times 10^{5}$
& $7\times 10^{3}$
& $1\times 10^{3}$
& $400$
& $200$
& $90$
      \\
$\frac{\mu_{eff}}{m_{soft}}$ 
& $K$
& 1
& 1
& 1
& 1
& 1
& 1
& 1
      \\
 
& $K+1$
& $2\times 10^{-8}$
& $7\times 10^{-6}$
& $1\times 10^{-4}$
& $8\times 10^{-4}$
& $3\times 10^{-3}$
& $6\times 10^{-3}$
& $1\times 10^{-2}$
      \\
\hline
& $0$
& 1
& 1
& 1
& 1
& 1
& 1
& 1
      \\
 
& $1$
& $2\times 10^{-8}$
& $7\times 10^{-6}$
& $1\times 10^{-4}$
& $8\times 10^{-4}$
& $3\times 10^{-3}$
& $6\times 10^{-3}$
& $1\times 10^{-2}$
      \\
& $2$
& $3\times 10^{-16}$
& $5\times 10^{-11}$
& $2\times 10^{-8}$
& $6\times 10^{-7}$
& $7\times 10^{-6}$
& $4\times 10^{-5}$
& $1\times 10^{-4}$
      \\
$\frac{m_{Q,L}}{\langle H_i\rangle} $ 
& $3$
& $6\times 10^{-24}$
& $3\times 10^{-16}$
& $2\times 10^{-12}$
& $5\times 10^{-10}$
& $2\times 10^{-8}$
& $2\times 10^{-7}$
& $2\times 10^{-6}$
      \\
& $4$
& $1\times 10^{-31}$
& $2\times 10^{-21}$
& $3\times 10^{-16}$
& $4\times 10^{-13}$
& $5\times 10^{-11}$
& $1\times 10^{-9}$
& $2\times 10^{-8}$
      \\
& $5$
& $2\times 10^{-39}$
& $2\times 10^{-26}$
& $5\times 10^{-20}$
& $3\times 10^{-16}$
& $1\times 10^{-13}$
& $9\times 10^{-12}$
& $2\times 10^{-10}$
      \\
\hline
\hline
\end{tabular}
\caption{Non-Renormalizable
MSSM mass terms via $\langle S \rangle$. For
$m_{soft}\sim 100$ GeV, $\MK\sim 3\times 10^{17}$ GeV.}
\label{masstable2}
\end{table}

The neutrino mass terms in (\ref{nonrenqlb}) offer various 
possibilities\footnote{Other applications of non-renormalizable operators 
to neutrino mass include \cite{nronu} and \cite{clseesaw}.}
for achieving small neutrino masses \cite{numass}, 
some not involving a traditional seesaw mechanism \cite{seesaw}.
Very light non-seesaw doublet neutrino Majorana masses are possible
via $W^{\rm (Maj)}_{L_iL_i}$ of the form,
\begin{equation}
m_{L_iL_i} \sim \frac{\langle H_2\rangle^2}{\MK} 
\left(\frac{m_{soft}}{\MK}\right)^{\frac{P^{''}_{L_iL_i}}{K+1}}
\sim   \langle H_2\rangle
       \left(\frac{m_{soft}}{\MK}\right)\times
\left(\frac{m_{soft}}{\MK}\right)^{\frac{P^{''}_{L_iL_i}}{K+1}} \ll 1
{\rm \phantom{a}eV} 
\label{neut1}
\end{equation}
The upper bound 
on neutrino masses from this term 
({\it i.e.}, the case of $P^{''}_{L_iL_i}= 0$) is 
around $10^{-4}$ eV 
(using $\langle H_2\rangle \sim m_{soft} = 100$ GeV
and $\MK = 3\times 10^{17}$ GeV), which is too small to be relevant
to dark matter or MSW conversions in the sun \cite{numass}.

If $W^{\rm (Maj)}_{L_iL_i}$ is not present,  
a superpotential term like 
$W^{\rm (Dir)}_{L_i {\nu}^c_i}$ can naturally yield heavier
physical Dirac neutrino masses of the form
\begin{equation}
m_{L_i {\nu}^c_i} \sim \langle H_2\rangle
\left(\frac{m_{soft}}{\MK}\right)^{\frac{P'_{L_i {\nu}^c_i}}{K+1}}.
\label{neut3}
\end{equation} 
For example, for $K=5$
the experimental neutrino upper mass limits given in Table
\ref{masstable1} allow 
$P'_{L_1 {\nu}^c_1}\ge 4$,
$P'_{L_2 {\nu}^c_2}\ge 3$, and
$P'_{L_3 {\nu}^c_3}\ge 2$.
Masses corresponding to $P'_{L_i {\nu}^c_i}= 4$ or $5$
($m_{L_i {\nu}^c_i} = 0.9$ eV or $10^{-2}$ eV, respectively)
are in the range interesting for solar and atmospheric 
neutrinos, oscillation experiments, and dark matter.

Neutrino singlets can acquire a Majorana mass
through $W^{\rm (Maj)}_{{\nu}^c_i}$,
\begin{equation}
m_{{\nu}^c_i {\nu}^c_i} \sim m_{soft}
\left(\frac{m_{soft}}{\MK}\right)^{\frac{\bar P_{{\nu}^c_i{\nu}^c_i}-K}{K+1}},
\label{neut2}
\end{equation}
which can be very large or small, depending on the sign of  
$\bar P_{{\nu}^c_i {\nu}^c_i}-K$. Laboratory and cosmological constraints
depend on the ${\nu}^c_i$ lifetimes (if it decays), cosmological
production and annihilation rates, and mixings with each other and with 
doublet neutrinos. These in turn depend on other couplings, such as  
$W^{\rm (Dir)}_{L_i {\nu}^c_i}$ or renormalizable couplings not associated
with the mass. Generally, however, the constraints are very weak due to
the absence of normal weak interactions, especially for heavy ${\nu}^c_i$
$({\bar P_{{\nu}^c_i{\nu}^c_i}\le K})$.

If both $W^{\rm (Dir)}_{L_i {\nu}^c_i}$  and
$W^{\rm (Maj)}_{{\nu}^c_i{\nu}^c_i}$ terms are present, 
the standard seesaw mechanism 
can produce light neutrinos 
via diagonalization of the mass matrix for eqs.~(\ref{neut3},\ref{neut2}).
The light mass eigenstate is
\begin{equation}
m^{\rm light}_{\rm seesaw} \sim
m_{L_i {\nu}^c_i}^2/ m_{{\nu}^c_i {\nu}^c_i} \sim m_{soft}
\left(\frac{m_{soft}}{\MK}\right)^{\frac{2P'_{L_i {\nu}^c_i}
+ K - \bar{P}_{{\nu}^c_i {\nu}^c_i}}{K+1}},
\label{neut4}
\end{equation}
while the heavy mass eigenstate is to first order
$m_{{\nu}^c_i{\nu}^c_i}$ as given by (\ref{neut2}). 
Various combinations of $K$, $P'_{L_i {\nu}^c_i}$ 
and $\bar P_{{\nu}^c_i {\nu}^c_i}$ produce
viable masses for three generations of light neutrinos.
For example, with $K=5$ and 
$P'_{L_i {\nu}^c_i}= P'_i = \{2,1\}$
for $i=1,\, 2$, respectively (the values of 
$K$ and $P'_{i=1,2}$
discussed above 
for the quarks and electrons), 
and with either $P'_{L_3{\nu}^c_3} = 1$ 
or $P'_{L_3{\nu}^c}= P'_{u_3} = 0$ 
(involving a renormalizable Dirac neutrino term), the light
eigenvalues of the three generations fall into the hierarchy of
$3\times 10^{-5}$ eV, $1\times 10^{-2}$ eV, 
and either $1\times 10^{-2}$ eV or $5$ eV for
$\bar{P}_{{\nu}^c_i {\nu}^c_i}= P'_{L_i {\nu}^c_i} + 1$. 
This range is again of interest for laboratory and non-accelerator experiments.

\section{Conclusions}
We have explored the nature of intermediate scale scenarios for
effective supergravity models as derived within a class of string vacua.
In particular, we explored  a class of string models which, along with
the SM gauge group and the MSSM particle content, contain massless SM
singlet(s) $S_i$. In addition, we assumed that
the effects of supersymmetry breaking are
parameterized by  soft mass parameters.

The necessary  condition for the intermediate mass scenario is the existence
of $D$-flat and $F$-flat directions  in the renormalizable part of the
$S_i$ sector.
In this case, the only renormalizable terms of the  potential are  due to 
the soft mass-square
parameters $m_i^2$.  If the  running of the soft mass parameters
is such that the effective  mass-square, along the flat direction,  becomes
negative at $\mu_{RAD}\gg M_Z$, the $S_i$'s 
acquire a non-zero VEV at an intermediate scale.
(Another possibility is that individual mass squares, 
but not the effective combination for the $D$-flat
direction, are negative. 
Then the VEV is of the order of the electroweak scale.)

Importantly, in a large number of string models, in particular
for a class of fermionic constructions,  there exist 
SM singlets $S_i$  with flat directions at the renormalizable level, which
 couple to additional exotic particles via Yukawa couplings  of ${\cal O}(1)$.
Such Yukawa couplings in turn ensure the radiative breaking, by
driving the soft $m_i^2$ parameters negative at $\mu_{RAD}\gg M_Z$.

For simplicity we confined the concrete analysis to
the case in which there is an additional $U(1)'$ symmetry, and two SM singlets
$S_{1,2}$ have opposite signs of the $U(1)'$ charges, thus ensuring
$D$-flatness for $|Q_1||S_1|^2=|Q_2||S_2|^2$
(similar results are expected for the case of a single standard model 
singlet and no additional $U(1)'$). In the analysis
of radiative breaking we considered three types  of Yukawa
couplings (of ${\cal O}(1)$) of $S_i$ to  the exotic particles 
and a range of the boundary conditions on soft mass
parameters at $\MS$.
For a large range of parameters we  obtained $\mu_{RAD}$
in the range  $10^5$ GeV to $10^{16}$ GeV (or at $\mu_{RAD}\sim
M_Z)$. 

In addition, we discussed the competition between the  effects of
the pure radiative breaking  ($\langle S\rangle \sim \mu_{RAD}$) and 
the stabilization of vacuum due to the non-renormalizable terms   in
the superpotential of the type $\hat{S}^{K+3}/{\cal M}^K$  
($\langle S\rangle \sim (m_{soft} {\cal M}^{K})^{1\over{K+1}}$). 
Non-renormalizable terms in the
superpotential  are generic (and calculable) in string models. For a class
of fermionic constructions $M\sim \MS$.  These terms are
dominant for $(m_{soft}{\cal M}^{K})^{1\over{K+1}}<\mu_{RAD}$.

In the case of the  pure radiative breaking,  the mass of the Higgs
field (and its fermionic partner) 
associated with non-zero VEV of $S$ is light and
of order $M_Z/(4\pi)$.
On the other hand, the breaking due to the
non-renormalizable terms implies a light Higgs field and the
supersymmetric partner both with the mass of 
order $M_Z$.

The non-renormalizable couplings of $S_i$'s to the MSSM particles in the 
superpotential in turn provide  a mechanism to obtain an effective  $\mu$
parameter and  the  masses for quarks and leptons. In the case of
the  pure
radiative breaking the  precise values of the $\mu$ parameter and  the
lepton-quark masses crucially depends on $\mu_{RAD}$. 
When the non-renormalizable terms dominate, 
these parameters assume specific values in terms of
$K$ and the order $P$ of the
non-renormalizable term by which  they are induced.
In particular, $\mu={\cal O}(m_{soft})$ for  $K=P$, thus providing a
phenomenologically acceptable value for the $\mu$ parameter.
(Another possibility is that in which the $U(1)'$ is broken at the 
electroweak scale and the effective $\mu$ is generated by a renormalizable 
term.)
We are able to obtain interesting hierarchies for the quark and lepton masses
for appropriate values of $P$. Also, small (non-seesaw) Dirac or Majorana
neutrino masses can be obtained, or the traditional seesaw mechanism
can be incorporated, depending on the nature of the non-renormalizable 
operators.

In conclusion, the string models provide an important framework in which the
intermediate scales can naturally occur and provide interesting implications for
the $\mu$ parameter and the fermion mass hierarchy of the MSSM sector.

\acknowledgments
This work was supported in part by U.S. Department of Energy Grant No. 
DOE-EY-76-02-3071. We thank P.~Steinhardt and I.~Zlatev for useful 
discussions.

\newpage
\appendix
\section{Renormalization Group Analysis}
If the standard model singlet $S_1$ (with $U(1)'$ charge $Q_1$) couples in the
superpotential
\begin{equation}
\label{super}
W=h \hat{S}_1 \hat{E}_1 \hat{E}_2,
\end{equation}
to a set of exotic fields $\hat{E}_{1,2}^{\alpha}$ (in general non-singlets under
the
standard model group; the index $\alpha$ is a multiplicity index not necessarily
associated with a gauge symmetry), the one-loop RGEs for couplings and soft masses
can be integrated analytically.

The RGE equations have the general 
form\footnote{The local (or global) symmetry 
associated with the multiplicity in $\alpha$ for the $E_{1,2}$ fields
permits us to write the Yukawa coupling $h$ and the soft masses of these
fields with no $\alpha$ indices.}
($t=\frac{1}{16\pi^2}\ln\frac{\mu}{M_{str}}$):
\begin{equation}
\frac{d g_a}{dt}=b_ag_a^3,\;\;\;\; \frac{d M_a}{dt}=2b_ag_a^2M_a,
\end{equation}
where the index $a$ runs over the different gauge group factors, with gauge
coupling $g_a$ and gaugino mass $M_a$, and
$b_a=\sum_RS(R_a) - 3 C(G_a)$.
The sum extends over chiral multiplets with $S(R_a)$ the Dynkin index of
the corresponding representation and $C(G_a)$ the quadratic Casimir invariant
of the adjoint representation. With the MSSM particle content, $b_3=-3$, $b_2=1$,
and $b_1=\frac{33}{5}$. In the case of two fundamental $SU(5)$ multiplets added
to the MSSM particle content, $b_3=-2$, $b_2=2$, and $b_1=\frac{38}{5}$.  
In writing these equations we are neglecting the possible kinetic mixing 
\cite{mix,mixdan}
between $U(1)_Y$ and $U(1)'$.

For the Yukawa coupling in eqn.~(\ref{super}):
\begin{equation}
\frac{dh}{dt}=(T+2) h^3 - h \sum_a r_a g_a^2,
\end{equation}
where
$T=\sum_{\alpha}\delta_{\alpha}^{\alpha},\;
r_a=2[C_a(S_1)+C_a(E_1)+C_a(E_2)].$  In Table V, we list the values of $T$ and
$C_a(E_i)$ for specific examples of $E_i$.

For the associated soft trilinear coupling:
\begin{equation}
\frac{dA}{dt}=2 (T+2) A h^2 - 2 \sum_a r_a g_a^2 M_a.
\end{equation}
Finally, for the soft masses of the scalar components of $S_1$ and
$D_{1,2}$

\begin{eqnarray}
\frac{dm_1^2}{dt}&=&2 T  h^2 \sigma^2 - 8 \sum_a g_a^2 C_a(S_1) M_a^2 + 
2 {\sum_a}' k_a^{-1}g_a^2 Q_a(S_1) {\mathrm Tr} [Q_a m^2],\\
\frac{d m_{E_{1,2}}^2}{dt}&=& 2 h^2 \sigma^2 - 8 \sum_a g_a^2 C_a(E_{1,2}) M_a^2
+ 2 {\sum_a}'k_a^{-1} g_a^2 Q_a(E_{1,2}) {\mathrm Tr} [Q_a m^2].
\end{eqnarray}

We used
$\sigma^2=m_1^2+m_{E_1}^2+m_{E_2}^2+A^2$,
the primed sum extends only to Abelian gauge group factors, and the $k_a$ are
normalization factors for the Abelian groups (e.g., $k_1=\frac{5}{3}$ in a GUT
normalization).

\begin{table}[tbp]
\begin{center}
\begin{tabular}{||c||c|c|c|c|c||}
$E_i\sim(SU(3),SU(2),U(1)_Y,U(1)')$&$T$&$C_3(E_i)$&$C_2(E_i)$&$C_1(E_i)$&$C_{1'}(E_i)
$\\
\hline
$D\sim(3,0,Y_D,Q_{D})$&3&$\frac{4}{3}$&0&$3Y^2_D/5$&$Q^2_{D}/k_{1'}$\\
$L\sim(0,2,Y_L,Q_{L})$&2&0&$\frac{3}{4}$&$3Y^2_L/5$&$Q^2_{L}/k_{1'}$\\
$S_i\sim(0,0,0,Q_{S_i})$&$N_p$&0&0&0&$Q^2_{S_i}/k_{1'}$
\end{tabular}
\end{center}
\caption{ \footnotesize Coefficients in RGEs for coupling of $\hat{S}_1$ to 
triplets, doublets, and $N_p$ pairs of identical MSSM singlets, 
via the superpotential $W=h\hat{S}_1\hat{E}_1\hat{E}_2$. For the numerical work
we chose $k_{1'}=\frac{5}{3}$.}
\end{table}

The solutions for this set of equations\footnote{Assuming universality of
the soft masses at the string scale, ${\mathrm Tr} [Q_a m^2]=0$ at all
scales for non-anomalous Abelian groups.} are:
\begin{eqnarray}
g_a^2(t)&=&\frac{g_0^2}{1-2b_ag^2_0t},\vspace{0.1cm}\\
M_a(t)&=&M_{1/2}\frac{g_a^2(t)}{g_0^2},\vspace{0.1cm}\\
h^2(t)&=&\frac{E(t) h^2_0}{1+ (T+2)h^2_0F(t)},
\vspace{0.1cm}\\
A(t)&=&A_0 \epsilon_f(t)+M_{1/2}\left[H_2(t)-
(T+2)h_0^2H_3(t)\epsilon_f(t)\right],\vspace{0.1cm}\\
m_1^2(t)&=&\left[1-3TR_f(t)\right]m^2_0-
TR_f(t)\epsilon_f(t)A^2_0-2TR_f(t)
\epsilon_f(t)\frac{H_3(t)}{F(t)}A_0M_{1/2}
\vspace{0.1cm}\nonumber\\
&+&M_{1/2}^2\left\{I_1(t)-TR_f(t)\frac{J(t)}{F(t)}+
T(T+2)\left[\frac{H_3(t)}{F(t)}\right]^2R_f^2(t)\right\},
\vspace{0.1cm}\\
m_{E_{1,2}}^2(t)&=&\left[1-3R_f(t)\right]m^2_0-
R_f(t)\epsilon_f(t)A^2_0-2R_f(t)
\epsilon_f(t)\frac{H_3(t)}{F(t)}A_0M_{1/2}
\vspace{0.1cm}\nonumber\\
&+&M_{1/2}^2\left\{I_{E_{1,2}}(t)-R_f(t)\frac{J(t)}{F(t)}+
(T+2)\left[\frac{H_3(t)}{F(t)}\right]^2R_f^2(t)\right\},
\end{eqnarray}
where
\begin{eqnarray}
E(t)&=&\prod_a[1-2b_ag_0^2t]^{r_a/b_a},\vspace{0.1cm}\\
F(t)&=&2\int_t^0E(t')dt',\vspace{0.1cm}\\
\epsilon_f(t)&=&\frac{1}{1+(T+2)h_0^2F(t)},\vspace{0.1cm}\\
R_f(t)&=&h_0^2F(t)\epsilon_f(t),\vspace{0.1cm}\\
H_2(t)&=&-2\sum_ar_ag^2_a(t)t,\vspace{0.1cm}\\
H_3(t)&=&-2t E(t) - F(t),\vspace{0.1cm}\\
I_k(t)&=&2\sum_aC_a(k)\frac{1}{b_a}\left[
1-\frac{1}{(1-2b_ag_0^2t)^2}\right]\vspace{0.1cm}\\
J(t)&=&2\int_t^0E(t')\left[H_2^2(t')+I_1(t')+I_{E_1}(t')+I_{E_2}(t')\right]dt'.
\end{eqnarray}
It is instructive to write some of these functions in terms of the
Yukawa coupling at its (pseudo) fixed point 
\begin{equation}
h_f^2(t)\equiv\frac{E(t)}{(T+2)F(t)}.
\end{equation}
This is the running Yukawa coupling for large $h_0$. 
It determines the
infrared fixed point at the electroweak scale, and when $h_0\sim 1$ it is expected 
that $h(t)$ approaches $h_f(t)$ for sufficiently large $-t$ (see Figure~1~(a)).
When $h(t)\simeq h_f(t)$ is a good approximation, many terms simplify in the
previous analytic solutions. In particular,
\begin{equation}
\epsilon_f(t)\equiv 1-\frac{h^2(t)}{h_f^2(t)}\rightarrow 0,\;\;
R_f(t)\equiv \frac{1}{(T+2)}\frac{h^2(t)}{h_f^2(t)}\rightarrow \frac{1}{(T+2)}.
\end{equation}

In addition to the exact analytical solutions presented above, it is useful to 
consider approximate analytical solutions to the RGEs.  
In this semi-analytic approach,
the running gauge couplings and gaugino masses are replaced by their 
average values,
\begin{eqnarray}
\label{semigcpl}
\bar{g}_{a}&=&\frac{1}{2}(g_{a}(M_{Z})+g_{0}),\\
\label{semigino}
\bar{M}_{a}&=&\frac{1}{2}(M_a(M_Z)+M_{1/2}).
\end{eqnarray}

With these approximations, the RGEs for the Yukawa coupling, soft trilinear
coupling, and the soft masses can be solved easily.  The Yukawa coupling has
the approximate solution
\begin{eqnarray}
\label{semiyuk}
h^2(t)=\frac{{\tilde{g}}^2}{1-X(t)},
\end{eqnarray}
in which
\begin{eqnarray}
\label{g2def}
{\tilde{g}}^2&=&\frac{1}{T+2}\sum_{a}r_a\bar{g}^2_{a},\\
\label{Xdef}
X(t)&=&X_0e^{2(T+2){\tilde{g}}^2t},\\
\label{X0def}
X_0&=&1-(1-\frac{{\tilde{g}}^{2}}{h^{0\,2}}).
\end{eqnarray}
The approximate solution for the trilinear coupling is given by
\begin{eqnarray}
\label{semiA}
A(t)=\frac{A_{0}X(1-X_0)}{X_{0}(1-X)}+
\tilde{m}_{\lambda}\frac{X}{1-X}\left[\frac{1}{X}
-\frac{1}{X_0}+\ln{\frac{X}{X_0}}\right],
\end{eqnarray}
where 
\begin{eqnarray}
\label{semiml}
\tilde{m}_{\lambda}=\frac{1}{(T+2){\tilde{g}}^2}\sum_{a}r_a\bar{g}^2_{a}\bar{M}_a,
\end{eqnarray}
and the other quantities are defined above.

If the $U(1)'$ factors are neglected, the soft scalar mass-squared parameters 
have the following approximate solutions:
\begin{eqnarray}
\label{semim1}
m^2_1&=&(1-\frac{T}{T+2})m^2_0+\frac{T}{T+2}\Sigma(t)+
2T{\tilde{g}}^2{\tilde{m}}^2\ln{\frac{X}{X_0}},\\
\label{semid1}
m^2_{E_{1,2}}&=&(1-\frac{1}{T+2})m^2_0+\frac{1}{T+2}\Sigma(t)+
(2{\tilde{g}}^2{\tilde{m}}^2
-8\sum_{a}C_a(E_{1,2})\bar{M}^2_a\bar{g}^2_{a})\ln{\frac{X}{X_0}},
\end{eqnarray}
in which 
\begin{eqnarray}
{\tilde{m}}^2&=&\frac{4}{(T+2){\tilde{g}}^2}
\sum_{a}C_a(E_{1,2})\bar{M}^2_a\bar{g}^2_{a},
\end{eqnarray}
and
\begin{eqnarray}
\label{semisum}
\Sigma(t)&=&({\tilde{m}}^2-{\tilde{m}_{\lambda}}^2)\frac{1-\frac{X}{X_0}}{1-X}
+\frac{X(1-X_0)}{X_0(1-X)}\Sigma_0-\frac{X(1-X_0)}{X_0}{(A_0
-\tilde{m}_{\lambda})}^2\frac{1-\frac{X}{X_0}}{{(1-X)}^2}\nonumber\\
&+&\frac{X(1-X_0)}{X_0{(1-X)}^2}2(A_0-\tilde{m}_{\lambda})\tilde{m}_{\lambda}
\ln{\frac{X}{X_0}}+\frac{X}{1-X}\ln{\frac{X}{X_0}}({\tilde{m}}^2+
{\tilde{m}_{\lambda}}^2\frac{1}{1-X}\ln{\frac{X}{X_0}}).
\end{eqnarray}
The semi-analytic solutions are valid in the limit of small initial gaugino
masses, such that the contribution of the gauginos to the evolution of the
trilinear coupling and the soft mass-squared parameters is small.
\end{document}